\documentclass[PAPER, atlasdraft=false, texlive=2023,cernpreprint,UKenglish, texmf, orcidlogo]{atlasdoc}
 
\usepackage[subfigure]{atlaspackage}
\usepackage{cancel}
\usepackage{atlasbiblatex}
 
\usepackage{array}
 
\usepackage{atlasphysics}

\graphicspath{{logos/}{figures/}}
 
\usepackage{ANA-TOPQ-2019-04-PAPER-defs}

\AtlasTitle{Search for flavour-changing neutral $tqH$ interactions
with $H\rightarrow \gamma\gamma$ in $pp$ collisions at $\sqrt{s}$ = 13
TeV using the ATLAS detector}
 
\AtlasVersion{2.3}
\AtlasAbstract{
A search for flavour-changing neutral interactions involving the top quark, the Higgs boson and an
up-type quark $q$ ($q = c, u$) is presented. The proton--proton collision data set used, with an
integrated luminosity of 139 fb$^{-1}$, was collected at $\sqrt{s} = 13$~TeV by the ATLAS experiment at
the Large Hadron Collider. Both the decay process $t \to qH$ in $t\bar{t}$
production and the production process $pp \to tH$,
with the Higgs boson decaying  into two photons, are investigated.  No significant excess is observed
and upper limits are set on the $t\rightarrow cH$ and the
$t\rightarrow uH$ branching ratios of $4.3\times 10^{-4}$ and $3.8\times 10^{-4}$, respectively,
at the 95\% confidence level, while the expected limits in the absence of signal are $4.7\times 10^{-4}$ and $3.9\times 10^{-4}$. Combining
this search with ATLAS searches in the $H \to \tau^+\tau^-$ and $H \to b\bar{b}$ final states yields observed
(expected) upper limits on the $t\to cH$ branching ratio of $5.8\times 10^{-4}\ (3.0\times 10^{-4})$ at the 95\% confidence level. The corresponding observed (expected)
upper limit on the $t\rightarrow uH$ branching ratio is $4.0 \times 10^{-4}\ (2.4 \times 10^{-4})$.}
 
\author{The ATLAS Collaboration}
 
\AtlasRefCode{TOPQ-2019-04}
 
\PreprintIdNumber{CERN-EP-2023-204}

\AtlasJournal{JHEP}
\AtlasJournalRef{JHEP 12 (2023) 195}
\AtlasDOI{10.1007/JHEP12(2023)195}

\hypersetup{pdftitle={ATLAS document},pdfauthor={The ATLAS Collaboration}}
 
\begin{document}
 
\maketitle

\tableofcontents
 
\section{Introduction}

% The next lines are included from the .//Sections/Intro.tex input file
 
Following the observation of the Higgs boson by the ATLAS~\cite{HIGG-2012-27} and the CMS~\cite{CMS-HIG-12-028} collaborations,
a comprehensive programme of measurements of its properties is underway. The search for flavour-changing
neutral current interactions (FCNC) between the Higgs boson, the top quark, and a charm or up quark is a part of the programme. Since the Higgs boson
is lighter than the top quark, such $tqH$ ($q = c, u$) interactions would manifest themselves in particular as FCNC top-quark decays, $t \to qH$.~\footnote{Throughout this paper the inclusion of charge-conjugate decay and production modes is implied.}
 
According to the Standard Model (SM), FCNC processes are forbidden at tree level and
very much suppressed at the one-loop level and higher orders due to the Glashow-Iliopoulos-Maiani (GIM) mechanism~\cite{Glashow:1970gm}.
Observations of FCNC decays of the top quark, which are extremely rare in the SM (with a branching ratio, ${\cal B}$, of about $4.2\times10^{-15}$
for $t \to cH$ and $3.7\times10^{-17}$ for $t \to uH$~\cite{Altmannshofer:2019ogm}), would constitute a clear signal of new physics.
 
In models beyond the SM, new flavour-changing mechanisms can contribute to the $tqH$ vertex, yielding effective couplings orders
of magnitude larger than those of the SM~\cite{AguilarSaavedra:2004wm}.
Examples of such extensions are the quark-singlet model~\cite{AguilarSaavedra:2002ns, delAguila:1998tp,
AguilarSaavedra:2002kr}, the two-Higgs-doublet model (2HDM) with or
without flavour violation~\cite{cheng:1987rs, Grzadkowski:1990sm, Luke:1993cy, Atwood:1996vj, Bejar:2000ub,Baum:2008, Branco:2011iw, Chen:2013, Atwood:2013},
the Minimal Supersymmetric Standard Model~\cite{Li:1993mg, deDivitiis:1997sh, Lopez:1997xv, Guasch:1999jp, Delepine:2004hr, Liu:2004qw, Cao:2007dk,Cao:2014udj},
Supersymmetry with R-parity violation~\cite{Yang:1997dk, Eilam:2001dh}, the Topcolour-assisted Technicolour model~\cite{Lu:2003yr},
models with warped extra dimensions~\cite{Agashe:2009di, Azatov:2009na} and the Littlest Higgs model with T-parity
conservation~\cite{2013:LHT}.
In composite Higgs boson models, FCNC may appear even with a single Higgs doublet~\cite{Agashe:2009di, Azatov:2014lha}.
For a review, see ref.~\cite{Agashe:2013hma}.
Among the studied potential branching ratio enhancements, the largest, ${\cal B}(t \to cH)$ of the order of $10^{-3}$, appears in the 2HDM with the ansatz
of Cheng and Sher~\cite{cheng:1987rs} where the off-diagonal Yukawa couplings, $\lambda_{tqH}$,
scale with the top- and charm- or up-quark masses, $m_t$ and $m_q$, as
$\lambda_{tqH}^{\text{CS}}=\sqrt{2m_qm_t}/v$ (where $v = 246\ \gev$ is the Higgs field vacuum expectation value).
Recently, Alves \textit{et al.}~\cite{Alves:2021pak} predicted a $t\to cH$ branching ratio between $1.8 \times 10^{-4}$ and $4.5 \times 10^{-4}$
within the context of a 2HDM model generating CP violation in the leptonic and hadronic sectors from a common origin.
 
Both the ATLAS and CMS collaborations have searched for $tqH$ couplings during \RunOne\ and \RunTwo of the LHC~\cite{HIGG-2013-09,TOPQ-2014-14,HIGG-2016-26,TOPQ-2017-15,TOPQ-2017-07,TOPQ-2019-17,HDBS-2020-16,CMS-TOP-13-017,CMS-TOP-17-003,CMS-TOP-20-007,CMS-TOP-19-002}.
In addition to the FCNC top-quark decay, $t \to qH$, the FCNC production process, $pp \to tH$, was for the first time
taken into account by CMS using the decay mode $H \to \bbbar$ with a data set corresponding to an integrated luminosity of 36~\ifb~\cite{CMS-TOP-17-003}, and
brought an improvement of about 20\% in the sensitivity to the $tuH$ coupling.
The CMS result~\cite{CMS-TOP-20-007} using the decay mode $H \to \gamma\gamma$ with 137 \ifb\ currently gives
the best limits. For the $tcH\ (tuH)$ couplings, 95\% confidence-level (CL) upper limits of $7.3\ (1.9) \times 10^{-4}$ were
observed while $5.1\ (3.1) \times 10^{-4}$ were expected.
Finally, with 139~\ifb\ of data at 13~\tev, ATLAS obtained 95\% CL upper limits of $9.4\ (6.9) \times 10^{-4}$
using $H \to \tau^+\tau^-$ decays~\cite{TOPQ-2019-17} and $12.0\ (7.7)\times 10^{-4}$
with $H \to \bbbar$ decays~\cite{HDBS-2020-16}.
 
The $tqH$ coupling induces both the $t \to qH$ decay and $pp \to tH$ production. Examples of Feynman diagrams for these processes are
shown in figure~\ref{fig:FeynDiag}.
\begin{figure}[h!]
\begin{center}
\hspace{-0.2 cm}
\subfigure[]{\includegraphics[width=.25\textwidth]{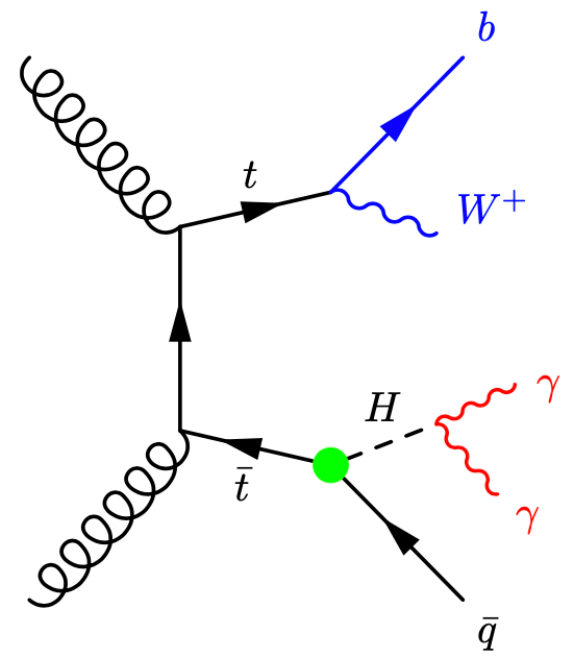}}\hspace{2cm}
\subfigure[]{\includegraphics[width=.25\textwidth]{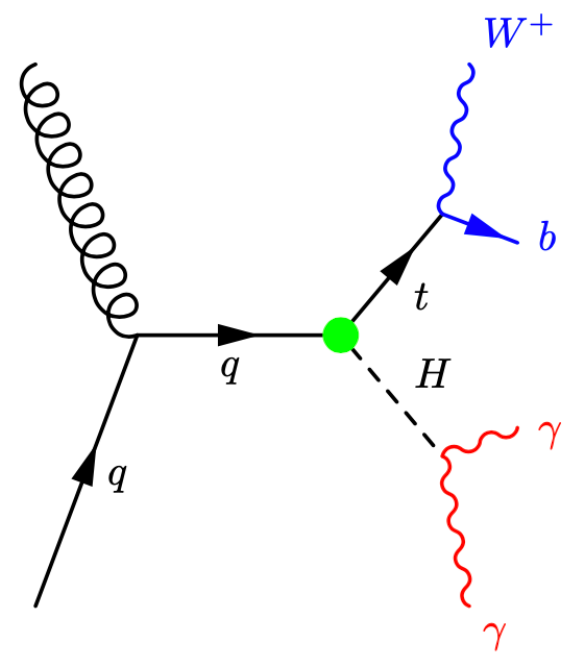}}\hspace{0.5cm}
\subfigure[]{\includegraphics[width=.25\textwidth]{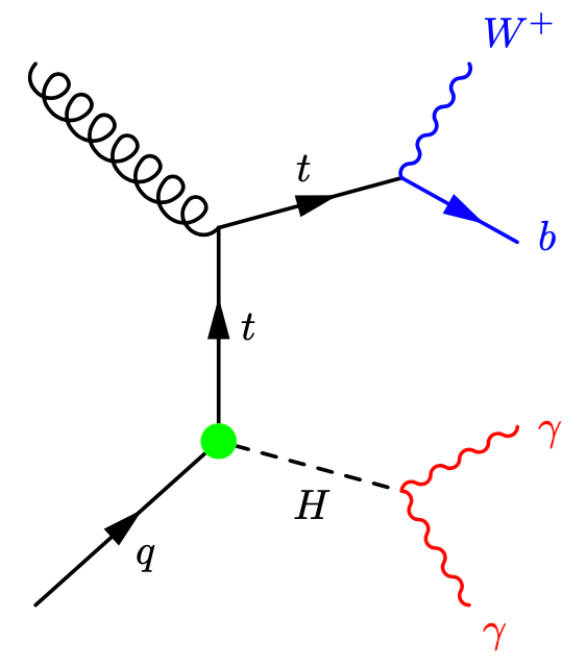}}
\end{center}
\vspace{-0.5cm}
\caption{Examples of leading-order Feynman diagrams for FCNC processes (a) in the top-quark decay and (b, c) in the associated production
of a top quark and a Higgs boson. The FCNC vertex is shown as a green filled circle.}
\label{fig:FeynDiag}
\end{figure}
In the SM effective field theory (SMEFT),
there are four FCNC operators contributing to the $tqH$ couplings at tree level. Using the notation
of ref.~\cite{Durieux:2014xla} and taking the same mass scale ($\Lambda$) for all operators, the
corresponding additional contribution to the SM Lagrangian is:
\begin{equation}
\delta{\cal L} =  \frac{-y_t^3}{\Lambda^2} (\varphi^{\dag}\varphi-v^2/2) \sum\limits_{i = 1,2} C_{u\varphi}^{i3}\overline{q_i}\tilde{\varphi}t+C_{u\varphi}^{3i}\overline{Q}\tilde{\varphi}u_i,
\end{equation}
where $y_t$ is the top-quark Yukawa coupling, $\varphi$ is the Higgs doublet ($\tilde{\varphi} = i\tau_2\varphi^{\ast}$), ($Q, q_i$) are the quark doublets
and ($t, u_i$) are the quark singlets; $Q$ and $t$ refer to the third generation and the index $i$ runs over the first
and second generations. The parameters $C^x_{u\varphi}\ (x = i3, 3i)$ are Wilson coefficients. Within SMEFT, a given branching ratio
of the process $t \to qH$ can be translated to a value of $C^x_{u\varphi}/\Lambda^2$. In a simple scenario
with a single operator, ${\cal B} = 10^{-3}$ and $\Lambda = 1$ \tev\ correspond to $C = 1.4$.
Limits on ${\cal B}$ can also be translated to limits on $\lambda_{tqH}$ via the relation
\begin{equation}
\lambda_{tqH} = (1.85\pm0.02)\times\sqrt{{\cal B}},
\label{eq:yukbr}
\end{equation}
where the mass of the light quark is neglected, the next-to-leading-order (NLO) estimations for the FCNC and SM widths
are used, and the values of the Wilson coefficients for the flavour-changing chromomagnetic operators
are assumed to be zero~\cite{HIGG-2013-09,Zhang:2013xya,Greljo:2014dka}. The uncertainty corresponds to missing
higher-order corrections. The $\lambda_{tqH}$ coupling corresponds to
the sum in quadrature of the couplings relative to the two possible chirality combinations of the quark fields,
$\lambda_{tqH} \equiv \sqrt{|\lambda_{t_{\tiny \text{L}}q_{\tiny \text{R}}}|^2+|\lambda_{q_{\tiny \text{L}}t_{\tiny \text{R}}}|^2}$.
 
In the search for the $tqH$ couplings in the decay of a top quark, only heavy-flavour tagging of reconstructed jets
can be used to disentangle the $c$ and $u$ flavours. The associated production of a single top quark and a Higgs boson
helps to lift the degeneracy between the $tcH$ and $tuH$ couplings from
the yields of the selected events. The up quark being a valence quark in the proton and the charm quark
a sea quark, the $ug \to tH$ yield is roughly seven times larger than the $cg \to tH$ one for equal couplings.
Moreover, the distributions of some quantities, such as the rapidity of the Higgs boson or the charge of the $W$ boson
from the top-quark decay, show clear differences between $cg \to tH$ and $ug \to tH$.
Both charm tagging and $tH$ production are used in this analysis.
 
The two final states, $tH$ and $tqH$, arising from the processes $pp \to tH$ and $pp \to \nttb \to t\qbar H$, respectively,
derive from the same coupling, and a single parameter of interest, namely the branching ratio for the decay $t \to qH$,
is extracted or constrained from the data, assuming only one coupling, either $tcH$ or $tuH$, is non-zero at a time.
The analysis presented here expands on methods already used in ref.~\cite{HIGG-2016-26}.
A profile likelihood fit of ${\cal B}$ is performed on several orthogonal regions that target events
with two photons from the Higgs boson decay, zero or one charged lepton, and are further subdivided on the basis of charm tagging and
top-quark reconstruction. In each category the background in the diphoton signal region is estimated in a data-driven approach by
a fit to the diphoton invariant mass distribution. Given the increase in the integrated
luminosity of close to a factor four, and the improved analysis, with a better categorisation and additional background rejection
using multivariate discriminants, a significant gain is expected compared with ref.~\cite{HIGG-2016-26}.

% End of text imported from the .//Sections/Intro.tex input file

\section{Detector, data set and Monte Carlo simulation}

% The next lines are included from the .//Sections/DetSim.tex input file
\subsection{ATLAS detector}
The ATLAS detector~\cite{PERF-2007-01} consists of an inner detector (ID) for tracking,
surrounded by a superconducting solenoid providing a 2\,T magnetic field,
electromagnetic and hadronic calorimeters, and a muon spectrometer. The inner detector provides tracking in the pseudorapidity
\footnote{ATLAS uses a right-handed coordinate system with its origin at the nominal interaction point
(IP) in the centre of the detector and the $z$-axis along the beam line. Observables labelled as transverse are projected onto the $x$--$y$ plane. The
$x$-axis points from the IP to the centre of the LHC ring, and the $y$-axis
points upwards. Cylindrical coordinates $(r, \phi)$ are used in the transverse
plane, $\phi$ being the azimuthal angle around the beam line. The pseudorapidity
is defined in terms of the polar angle $\theta$ as $\eta = -\ln\tan(\theta/2)$. The angular distance $\Delta R$ is defined as
$\Delta R \equiv \sqrt{(\Delta \eta)^2 + (\Delta \phi)^2}$. The transverse energy is $\et = E/\cosh(\eta)$.}
region $|\eta| < 2.5$ and consists of a silicon pixel detector, including the insertable B-layer~\cite{ATLAS-TDR-19,PIX-2018-001}
installed before \RunTwo, and a microstrip detector inside a transition radiation tracker that covers $|\eta| < 2.0$.
The electromagnetic calorimeter, a lead/liquid-argon sampling device with accordion geometry, is divided into
one barrel ($|\eta|<1.475$) and two endcap ($1.375<|\eta|<3.2$) sections.
Longitudinally, it is divided into three layers. While most of the energy is deposited in the second layer, the first layer, referred to as
the strip layer, has fine segmentation in the regions $|\eta|<1.4$ and $1.5<|\eta|<2.4$ to help the separation
of photons from neutral hadrons and to allow shower
directions to be measured. In the range of $\lvert\eta\rvert < 1.8$, a
presampler layer allows the energy to be corrected for losses
upstream of the calorimeter. The barrel ($|\eta|<1.7$) hadronic calorimeter consists
of steel and scintillator tiles, while the endcap sections ($1.5<|\eta|<3.2$)
are composed of copper and liquid argon. The forward calorimeter
($3.1<|\eta|<4.9$) uses copper and tungsten as absorber with liquid argon as active material.
The muon spectrometer consists of precision ($|\eta|<2.7$) and trigger ($|\eta|<2.4$) chambers equipping a toroidal magnet system
which surrounds the hadronic calorimeter. The field integral of the toroid magnets ranges between 2.0 and 6.0~Tm across most of the detector.
 
A two-level trigger system is used to select events. The first-level trigger is implemented in hardware and
uses a subset of the detector information to accept events at a rate below 100~kHz.
This is followed by a software-based trigger that reduces the accepted event rate to 1~kHz on average,
depending on the data-taking conditions~\cite{TRIG-2016-01}.
 
An extensive software suite~\cite{ATL-SOFT-PUB-2021-001} is used in the reconstruction and analysis of real and simulated data,
in detector operations, and in the trigger and data acquisition systems of the experiment.
 
\subsection{Data set}
\par This analysis uses the full proton--proton collision data set recorded by the ATLAS detector from 2015 to 2018 (\RunTwo) with the LHC operating at a centre-of-mass energy of $\sqrt{s} = 13$~\tev\ and a bunch spacing of 25~ns. After application of data-quality requirements~\cite{DAPR-2018-01}, the integrated luminosity amounts to 139~\ifb, with a relative uncertainty of 1.7\%~\cite{ATLAS-CONF-2019-021}, obtained using the LUCID-2 detector~\cite{LUCID2} for the primary luminosity measurements.
The data were recorded with instantaneous luminosities up to $1.9\times 10^{34}\,\text{cm}^{-2}\text{s}^{-1}$.
The mean number of interactions per bunch crossing, $\mu$, ranged from an average of 13 in 2015 to 38 in 2017, with a global average of about 34.
The inelastic collisions that occur in addition to the hard interaction produce mainly particles with low transverse momenta that form the \pileup background.
 
\subsection{Simulation samples}
\label{sec:MC}
Samples of simulated Monte Carlo (MC) events were produced to model the different signal and background
processes.
 
Two sets of signal samples, corresponding to the $pp \to \ttbar \to bW\bar{q}H$ and $pp \to tH \to bWH$
processes, were produced at NLO in QCD\@. The \ttbar production
was simulated using \POWHEGBOX[v2]~\cite{Frixione:2007nw} and the $tH$ production
with \MGNLO[2.6.0]~\cite{Alwall:2014hca} and the TopFCNC Universal FeynRules Output (UFO)~\cite{Degrande:2011ua} model~\cite{Degrande:2014tta}.
The top quarks are decayed by \MADSPIN~\cite{Artoisenet:2012st} using the TopFCNC UFO model, while the Higgs
boson decay into two photons is simulated in \PYTHIA~8.2~\cite{Sjostrand:2014zea}.
Both simulations use the \NNPDF[3.0nlo]~\cite{Ball:2014uwa} parton distribution functions (PDF) for the matrix element and are interfaced to
\PYTHIA~8.2 with the A14 tune~\cite{ATL-PHYS-PUB-2014-021}
for the parton shower, hadronisation and underlying event using the \NNPDF[2.3lo]~\cite{Ball:2012cx} PDF\@.
The top quark mass is set to 172.5~\gev, and the Higgs boson mass, \mH, is set to 125~\gev.
 
For \ttbar production, two MC samples with one top quark decaying into a charm quark
and a Higgs boson were produced. The two samples correspond to the leptonic and the hadronic decays of
the $W$ boson. The leptonic decays of the $W$ boson include all three lepton flavours
($W \to \ell\nu, \ell = e, \mu, \tau$). Equivalent samples with $t \to uH$ are also used.
The nominal renormalisation and factorisation scales in the \ttbar signal sample are chosen
to be equal and given by $\muF = \muR = \sqrt{m_t^2 + \pt^2}$, where \pt is the transverse
momentum of the top quark, and the $h_{\mathrm{damp}}$ value~\footnote{The $h_{\mathrm{damp}}$ parameter controls the transverse momentum of the first additional emission beyond the leading-order Feynman diagram in the parton shower and therefore
regulates the high-\pt emission against which the \ttbar system recoils.} is set equal to 1.5$m_t$~\cite{ATL-PHYS-PUB-2018-009}.
The cross-section for the $pp \to \ttbar \to bW\bar{q}H, H \to \gamma\gamma$ process is
$\sigma = 2\sigma_{\ttbar}{\cal B}_{\gamma\gamma}{\cal B}(1-{\cal B})$, where $\sigma_{\ttbar}$ is the
\ttbar production cross-section and ${\cal B}_{\gamma\gamma}$ the branching ratio for the $H \to \gamma\gamma$
decay. For these two quantities, the following SM values are used: $\sigma_{\ttbar} = 832\pm51$ pb, calculated at next-to-next-to-leading
order in QCD including resummation of next-to-next-to-leading logarithmic soft gluon terms with \TOPpp[2.0]
(see~\cite{Czakon:2011xx} and references therein), and ${\cal B}_{\gamma\gamma} = (2.27\pm 0.07) \times 10^{-3} $~\cite{deFlorian:2016spz}.
For a branching ratio for the FCNC top-quark decay of ${\cal B} = 10^{-3}$, just below the ATLAS combined
limit using 36 \ifb\ of data collected at $\sqrt{s} = 13$ \tev~\cite{TOPQ-2017-07}, the cross-section is
$3.77\pm0.25$ fb.
 
For the $tH$ signal simulation, the nominal renormalisation and factorisation scales are chosen to be equal and given by
$\muF = \muR = m_t + \mH$. The interference between double-resonant top production and single-top production
at NLO is neglected. Eight MC samples were generated, where the top quark decays into $bW$; for a given
quark flavour ($u$ or $c$) and $W$ decay ($W \to \ell\nu$ or \nqqbp) two samples were considered according to
the chirality combination of the involved quark fields.
It was observed that the chirality choice has a negligible impact on the kinematic distributions. The samples are
therefore combined in the following. The cross-sections for the $pp \to bWH, H \to \gamma\gamma$ process
are $1.61\pm0.13$ fb and $0.24\pm0.02$ fb for $q = u$ and $c$, respectively, for a single FCNC operator with
$C = 1.4$ and $\Lambda = 1$ \tev, see refs.~\cite{Degrande:2014tta,Durieux:2014xla}.
 
The contributions from the known Higgs boson production mechanisms were simulated for each of the five main SM modes:
gluon--gluon fusion (ggF), vector boson fusion (VBF), associated $WH$, $ZH$ and $\nttb H$ production,
and the rare processes \nbbb$H$, $tHq$ and $tWH$. The cross-sections given in ref.~\cite{deFlorian:2016spz} are used for normalisation,
except for ggF, for which the next-to-next-to-next-to-leading-order cross-section is used.
An event sample, labelled \Syyj\ in the following, is used as a benchmark sample for non-resonant diphoton production with a fully hadronic final state.
It was generated with \SHERPA 2.2.4~\cite{Bothmann:2019yzt} with up to one parton at NLO and up to three partons at LO, using the \NNPDF[3.0nnlo] PDF and the dedicated set of tuned parton-shower parameters developed by the \SHERPA authors. The (sub-)leading
photon \ET is required to be above (18) 20 \gev\ and the diphoton invariant mass is required to be in the range [90, 175]~\gev.
Other non-resonant background processes (in particular from \ttbar + $\gamma$ production) have also been simulated. More details are given in table~\ref{tab:simuSM}.
\begin{table}[h!]
\begin{center}
\caption{Summary of the background MC samples. For Higgs boson production, the $H \to \gamma\gamma$ branching ratio is included.
The non-resonant background samples are normalised according to the cross-sections provided by the generators, except the $Z \to ee$ sample,
which is normalised using a control data sample.
}
\label{tab:simuSM}
\footnotesize
\vspace{-0.2cm}
\begin{tabular}{lccccc}
\hline
Process          &  Generator & Showering & PDF set & Parameter tune  & cross-section (fb) \\
\hline
ggF           & \POWHEGBOX NNLOPS~\cite{Alioli:2008tz,Hamilton:2013fea}  & \PYTHIA 8.2 & \PDFforLHC[15] & \AZNLO~\cite{STDM-2012-23} & 110 \\
VBF           & \POWHEGBOX~\cite{Nason:2009ai} & \PYTHIA 8.2 & \PDFforLHC[15] & \AZNLO & 8.6 \\
$WH$        & \POWHEGBOX~\cite{Luisoni:2013kna} & \PYTHIA 8.2 & \PDFforLHC[15] & \AZNLO & 3.1 \\
$ZH$        & \POWHEGBOX~\cite{Luisoni:2013kna} & \PYTHIA 8.2 & \PDFforLHC[15] & \AZNLO & 2.0 \\
$\nttb H$ & \POWHEGBOX~\cite{Hartanto:2015uka} & \PYTHIA 8.2 &  \NNPDF[3.0nlo] & A14 & 1.2 \\
\nbbb$H$ & \POWHEGBOX & \PYTHIA 8.2 & \NNPDF[3.0nlo] & A14 & 1.1 \\
$tHjb$      & \MGNLO & \PYTHIA 8.2 & \NNPDF[3.0nlo] & A14 & 0.17 \\
$tWH$      & \MGNLO & \PYTHIA 8.2 & \NNPDF[3.0nlo] & A14 & 0.03 \\
\hline
$\gamma\gamma$ + jets & \SHERPA & \SHERPA  & \NNPDF[3.0nnlo] & -- & $51.8\times10^{3}$ \\
\nttb$\gamma$ & \MADGRAPH[5] & \PYTHIA 8.2 & \NNPDF[2.3lo] & A14 & $4.6\times10^{3}$  \\
$V \gamma\gamma, V = W,Z$ & \SHERPA & \SHERPA & \NNPDF[3.0nnlo] & -- & 236 \\
$Z \to ee$ & \POWHEGBOX & \PYTHIA 8.1 & \CT[10] & AZNLO & $2\times10^{6}$ \\
$tW\gamma$ & \MADGRAPH[5] & \PYTHIA 8.2 & \NNPDF[3.0nlo] & A14 & 533 \\
$tq\gamma$ & \MGNLO& \PYTHIA 8.2 & \NNPDF[3.0nlo] & A14 & 1139 \\
\hline
\end{tabular}
\end{center}
\end{table}
 
The stable particles, defined as particles with a lifetime larger than 10~ps, are passed through a full detector
simulation~\cite{SOFT-2010-01} based on \GEANT~\cite{Agostinelli:2002hh,Geant4PhysicsLists} for
the resonant background and $V\gamma\gamma$ processes. For other samples, a faster version of the simulation
was used, that relies on a parameterisation for the response of the calorimeters and on \GEANT for the other
components of the detector~\cite{SOFT-2010-01}. The resulting ``particle hits'' in the active detector material are later transformed
into detector signals during digitisation. \Pileup is modelled with simulated minimum-bias events generated with
\PYTHIA 8.186~\cite{Sjostrand:2007gs} using the \NNPDF[2.3lo] set of parton distribution functions and the A3 set of tuned parameters~\cite{ATL-PHYS-PUB-2016-017}.
The number of  events overlaid onto the hard-scattering events during digitisation is randomly chosen
to reproduce the distribution of $\mu$ observed in data. The effects of \pileup events occurring in nearby
bunch crossings (out-of-time \pileup) are also modelled.

% End of text imported from the .//Sections/DetSim.tex input file

\section{Event reconstruction and selection}

% The next lines are included from the .//Sections/EvtRecoSel.tex input file
Selected events must contain two isolated, high-\pt photons to tag the Higgs boson decay.
Additional objects, jets, leptons (electrons or muons) and missing transverse momentum, are requested
as signatures of top-quark decays.
The reconstruction and selection of all these objects are described below. A prerequisite is that at least one primary vertex is reconstructed in the event. In the general case of several reconstructed vertices, the $\gamma\gamma$ vertex is used~\cite{HIGG-2013-08}. It is determined taking into account the direction of each photon
(determined by exploiting the longitudinal segmentation of the calorimeter), \pT balance using charged-particle tracks from a given vertex and the two photons, and constraints from the longitudinal size of the luminous region. This vertex is used to correct the photon's four momenta and to construct the track-based isolation, jets, flavour tagging and \met.
 
\subsection{Photon reconstruction and identification}
The photon reconstruction~\cite{EGAM-2018-01} is seeded by clusters of energy deposits in the
calorimeter, formed using a dynamical, topological cell-clustering algorithm~\cite{PERF-2014-07}.
Clusters are accepted in the pseudorapidity region $|\eta| < 2.37$, with the exception of the transition region [1.37, 1.52],
where dead material affects both the identification and the energy measurement.
Misidentification of electrons as photon candidates is suppressed by checking
if a calorimeter deposit reconstructed as a photon has a matching track compatible with the primary vertex.
 
\noindent
Clusters without any matching track in the ID are classified as unconverted photon candidates.
Clusters with a matching conversion reconstructed from one or two tracks are classified as converted photon candidates.
The photon identification efficiency depends on the photon's transverse momentum, its pseudorapidity and whether it is classified as converted or unconverted~\cite{EGAM-2018-01}.
First a \textit{loose} identification criterion is required, which is based on requirements on their shower shape.
The two highest-\pT photon candidates, which must fulfil $\pT>25~\GeV$, are the main objects
for the analysis and are used to choose the primary vertex (see above).
After this preselection, the photons are further required to satisfy a \textit{tight} identification criterion to suppress fake photon candidates.

\noindent
The photon candidates are also required to satisfy \textit{loose} isolation criteria. The track-based isolation (sum of \pt of tracks in a cone of
$\Delta R=0.2$ around the photon candidate) must be smaller than $0.05\times\pT^{\gamma}$, where $\pT^{\gamma}$ is the photon's transverse momentum, and the
calorimeter-based isolation (sum of the transverse energy of topological clusters in a cone of $\Delta R=0.2$, corrected for
\pileup and photon energy leakage) must be smaller than $0.065\times\pT^{\gamma}$.
The dependence of the isolation efficiency on the event topology was assessed from a simulation-based study, giving
an efficiency (per diphoton event) of 89.5\% for the ggF production process and 82.1\% for the \nttb$H$ final state.
The track- and calorimeter-based isolation distributions are in good agreement between data and simulation. A scale factor is used to correct
for observed small differences. The results of the analysis are extracted using the sample of events where both
photons satisfy the \textit{tight} identification criterion and are isolated. An orthogonal sample, in which one
of the two photons does not pass the \textit{tight} identification or is not isolated, is used as a control sample at various stages of the analysis.
 
\noindent
The photon energy is determined in four steps using a combination of simulation-based
and data-driven calibration factors~\cite{EGAM-2018-01,PERF-2017-03}.
The data-driven calibration factors used to set the absolute energy scale are determined from $Z \to ee$ events. The photon energy resolution in simulation is corrected to match the resolution in data. This correction is derived simultaneously with the energy calibration factors using $Z\to ee$ events by adjusting the electron energy resolution such that the width of the reconstructed $Z$-boson peak in simulation matches the width observed in data.
 
\subsection{Reconstruction and selection of leptons, light- and heavy-flavour jets and missing transverse momentum}
\label{subsec:otherobj}
 
Electrons and muons are used in the leptonic selection. Electrons are reconstructed from energy clusters in the calorimeter associated with an ID track~\cite{EGAM-2018-01}.
Muon candidates are built from tracks reconstructed in the muon chambers~\cite{PERF-2015-10,MUON-2018-03}. A matching of these tracks to ID tracks is required in the region
$|\eta| < 2.5$. Muons are required to meet the conditions $|\eta| < 2.7$ and $\pT>10~\GeV$; for electrons the transverse momentum threshold is raised to  $\pT = 15~\GeV$, to remove fake electron candidates, which are more abundant at low $\pt$. Additionally, electrons must satisfy $|\eta|<2.47$, excluding the transition region. Both the electrons and muons must satisfy \emph{medium} identification and \emph{loose} isolation requirements~\cite{EGAM-2018-01,PERF-2015-10,MUON-2018-03}.
 
Jets are reconstructed using the \antikt algorithm~\cite{Cacciari:2008gp,Fastjet} with the radius parameter $R = 0.4$ and are required to have a rapidity $|y|<4.4$ and transverse momentum $\pT>25~\GeV$.
The objects used to form jets come from a particle-flow algorithm which combines information from the tracker and the calorimeters~\cite{PERF-2015-09}. To suppress \pileup jets, the \textit{tight}
working point of the \textit{jet vertex tagger} (JVT)~\cite{ATLAS-CONF-2014-018} is used (for jets with $\pt < 60\ \gev$). Jets with $|\eta|>2.5$ are also required to satisfy
the \textit{forward jet vertex tagger} (fJVT)~\cite{PERF-2016-06}. To limit further \pileup effects, the jet-\pT threshold is set to 30~\GeV.
 
Central jets ($|\eta|<2.4$) that are identified as originating from a \bquark are labelled as \bjets. The DL1r tagger~\cite{FTAG-2019-07} is used, with a 77\% efficiency working point, corresponding to a rejection of $\sim 250$ ($\sim 5.5$) for jets with \pt in the range [40,~60]~\gev originating from $u$-, $d$-, and $s$-quarks or gluons ($c$-quarks) in simulated $pp \to \ttbar$ events. The \pt calibration of \bjets is adjusted after they have been identified as such.~\footnote{The four-momentum of the highest-\pt muon found within a cone of
radius $R = \min(0.4,0.04 + 10\ \gev/\pt)$ around the jet axis is added to that of the jet, and a residual correction is applied to equalise the response to jets with leptonic or hadronic decays
of heavy-flavour hadrons.} A dedicated charm tagger has been optimised and calibrated for this analysis.
It uses the \bjet, \cjet and light-flavour jet probabilities ($p_b, p_c$ and $p_{\textrm{light}}$) from the DL1r tagger, and combines
them into a final discriminant defined as DL1r$_c = \ln\left(\frac{p_c}{f_b \cdot p_b + (1-f_b) \cdot p_{\textrm{light}}}\right)$.
The $b$-fraction parameter, $f_b$, and the threshold, DL1r$_c^{\text{thr}}$, for a jet to be \ctagged were optimised, leading to the choices $f_b = 0.2$ and DL1r$_c^{\text{thr}} = 1$.
The average efficiency of the charm tagging of jets originating from \cquarks in the \ttbar signal sample is 38\%.
The corresponding efficiency to tag jets originating from \bquarks (light-quarks or gluons, $\tau$-leptons) is about $15\%$ ($3\%, 31\%$).
Efficiency measurements for this tagger are performed simultaneously for all jet flavours ($b, c$ and light)
using semileptonic and dileptonic \ttbar events, and scale factors to be applied to jets in simulation have been derived. These scale factors are compatible with
unity for jets originating from \bquarks and slightly smaller (larger) than unity for jets from \cquarks ($u$-, $d$- or $s$-quarks or gluons).
The uncertainties in the scale factors for \bquark jets are about $5\%$.
The uncertainties in the scale factors for the \cquark jets and jets originating from light quarks or gluons
range from 5\% to 15\% being slightly higher for transverse momenta below 40~\gev.
 
In case of overlap between reconstructed particles, a removal is performed keeping, in order of priority, photons, then leptons, and finally jets.
Leptons or jets within a cone of radius $\Delta R = 0.4$ around photon candidates are removed first; then jets within $\Delta R = 0.2$ of electrons are removed; at last, leptons
within $\Delta R = 0.4$ of the remaining jets are removed.
 
\noindent
The missing transverse energy, \met, is computed as the negative sum of the transverse momenta of the reconstructed photons, electrons, muons and jets, plus a ``soft term'' reconstructed from all tracks not associated with any of the previous objects~\cite{PERF-2016-07}. Only tracks originating from the diphoton primary vertex are considered.
 
As already mentioned above, and more generally, differences between data and simulation are corrected for by using scale factors, applied as weights to MC events.
 
\subsection{Event preselection}
\noindent
Events were selected with a diphoton trigger requiring at least two candidate photons with \et greater than 35~\gev\ and 25~\gev, respectively.
Both photons are required to fulfil the \emph{loose} identification requirements for the 2015 and 2016 data sets, while \emph{medium} criteria are required for the 2017 and 2018 ones to cope with the larger instantaneous luminosity. The requirements are based on the energy leakage in the hadronic calorimeter and on the shower shape in the second and first two  layers of the electromagnetic calorimeter for the \emph{loose} and \emph{medium} criterion, respectively~\cite{TRIG-2018-05,TDAQ-2019-01}. The trigger selections are estimated to be fully efficient for photons satisfying the offline selection criteria discussed above and matched to the photons identified by the trigger.
 
The selection of  candidate events starts by
applying a tight diphoton selection: at least two photons sa\-tisfying the \emph{tight} identification criteria,
with \emph{loose} calorimeter-based and track-based isolation, ${\pt > 40\ \gev}$ (30~\gev) for the leading (sub-leading) photon candidate, and a diphoton invariant mass between 100~\gev\ and 160~\gev. Events without
identified lepton (electron or muon) enter the hadronic selection; those with exactly one lepton enter the leptonic selection. Events with two or more identified leptons are rejected.
Jets are ordered by decreasing value of \pt, and up to five jets are considered.
The various steps of the hadronic and leptonic selections are described in sections~\ref{subsec:FullHadronic} and~\ref{subsec:Leptonic}, respectively, and
a summary is given in diagrammatic form in section~\ref{subsec:summary}.
 
\subsection{Hadronic selection}
\label{subsec:FullHadronic}
 
The hadronic selection targets the processes $\nttb \to bW(\nqqbp)\bar{q}H(\gamma\gamma)$ and $tH\to bW(\nqqbp)H(\gamma\gamma)$, for which at least four and three jets in the final state are required, respectively. Selected events are classified in categories of decreasing purity, based on kinematic constraints and flavour tagging.
 
\subsubsection{Categories}
Addressing first the \nttb\ production, the reconstruction of the neutral current top-quark decay (called Top1 in the following) and the SM top-quark decay (called Top2) is described below. For events with four (five or more) jets, in total four (20) combinations are formed. Each combination has 1+3 jets, of which the first one is associated with the two photons in view of forming the Top1 combination, while the group of three is candidate to form Top2.  All considered jets that satisfy $152\ \gev < \mggj < 190\ \gev$ are subject to \ctag (and not to \btag). All other jets are subject to \btag. There must be exactly one \btagged jet among them. Figure~\ref{hadronic:fig:mtop12_hada} shows the \mggj\ invariant mass distribution, before the selection on this variable. Figure~\ref{hadronic:fig:mtop12_hadb} shows the invariant mass of the three other jets (among which one is \btagged) for those combinations passing the Top1 condition.
The Top2 condition is met if the 3-jet invariant mass satisfies  $120\ \gev < \mjjj < 220\ \gev$. The signal distribution corresponds to the $tcH$ coupling with an arbitrary $t\to cH$ branching ratio of ${\cal B}=2\%$. The background is the sum of the $tX\gamma$ ($X = t, W, q$) and $V\gamma\gamma$ contributions as described in section~\ref{sec:MC}, normalised to the cross-sections as given by the MC generators, together with the dominant \Syyj\ contribution. An additional $Z \to ee$ contribution, relevant for the leptonic analysis, is also included.
\par Based on the criteria above, events accepted at this stage can fall into one or more of the following four categories:
\begin{itemize}
\item had-\uA : at least one combination satisfying the Top1 and the Top2 mass requirements,
for which $j_0$, the jet forming the Top1 combination with the diphoton system, is \ctagged;
\item had-\uB : same as \uA\ except that the charm tagging condition is not met;
\item had-\dA : same as \uA\ except that the Top2 mass requirement is not met. The combination which includes the
\btagged jet and whose invariant mass is closest to 170~\gev\ is retained;
\item had-\dB : same as \dA\ except that the charm tagging condition is not met.
\end{itemize}
If an event falls into more than one of the categories, the one with the highest rank ({\uA\ $>$ \uB\ $>$ \dA\ $>$ \dB}) is kept.
\begin{figure}[h!]
\begin{center}
\hspace{-0.2 cm}
\subfigure[]{\includegraphics[width=.5\textwidth]{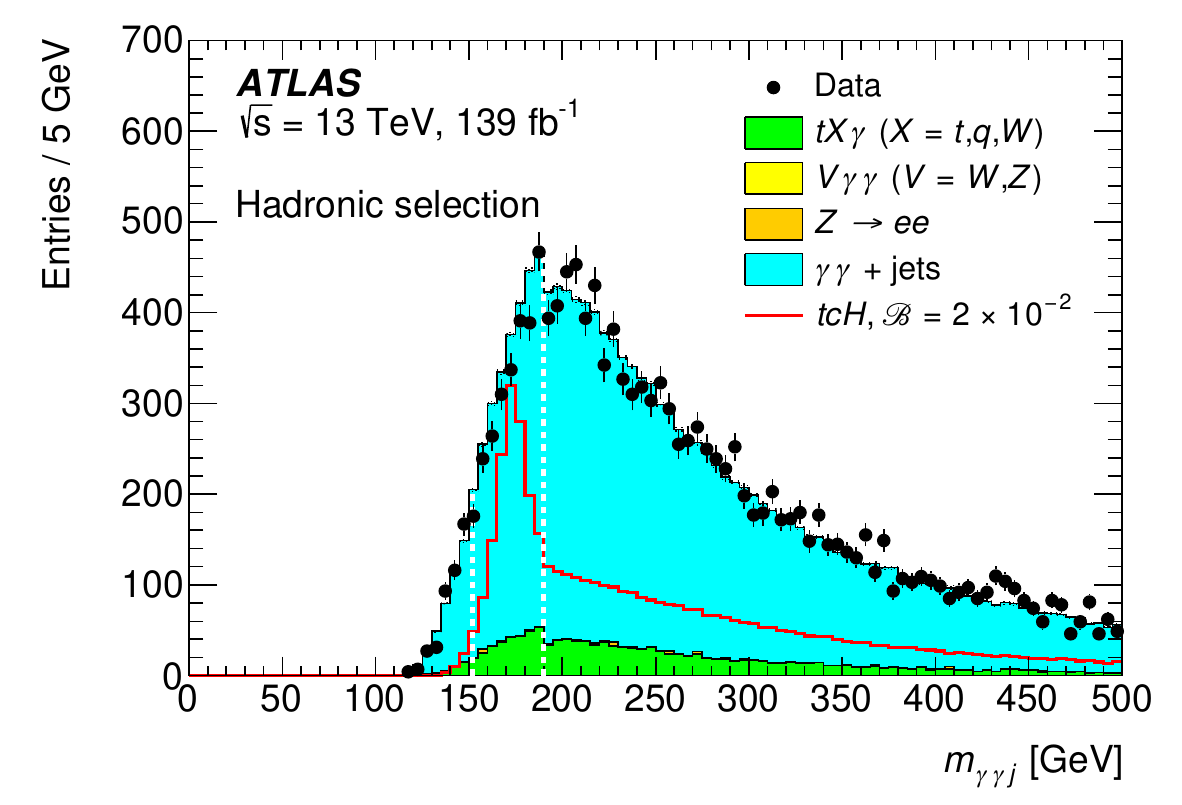}\label{hadronic:fig:mtop12_hada}}
\subfigure[]{\includegraphics[width=.5\textwidth]{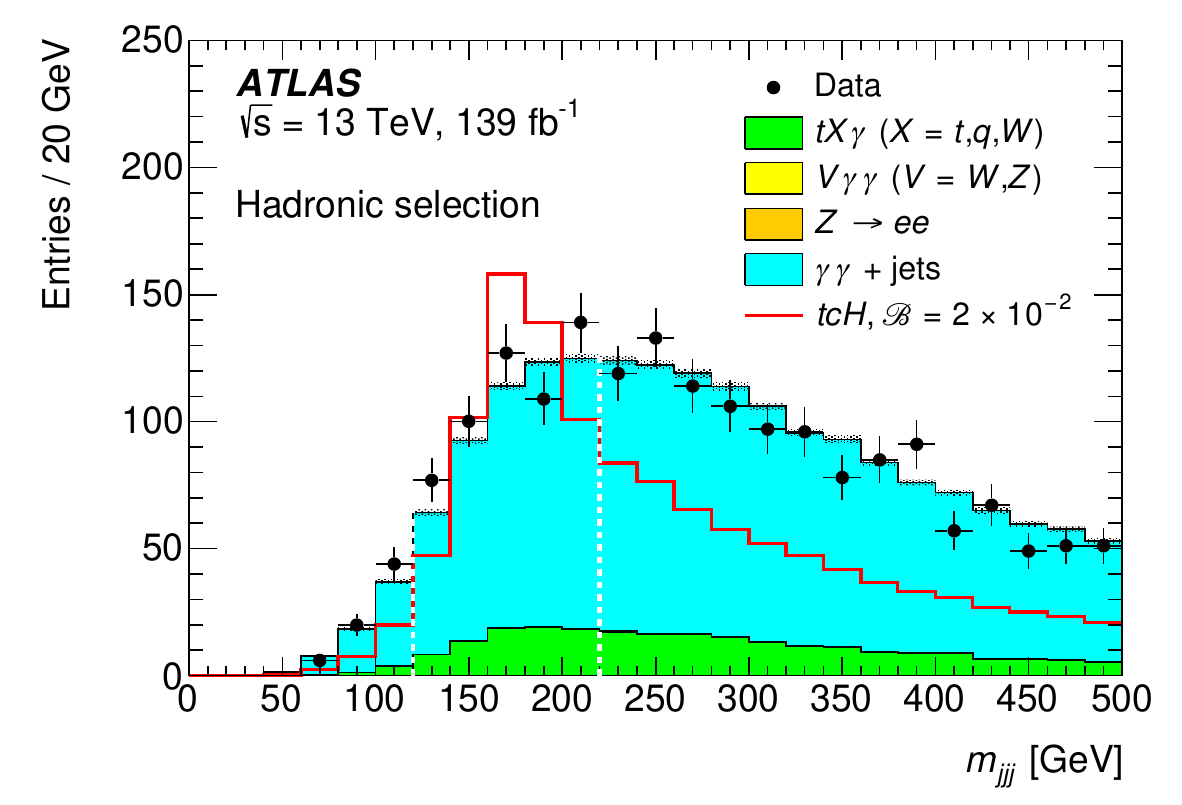}\label{hadronic:fig:mtop12_hadb}}
\end{center}
\vspace{-0.5cm}
\caption{Distributions of the invariant mass of (a) the two photons and one jet, when there is one \btagged jet among the three other jets which will be also tested against the Top2  mass condition
and (b) the three jets (among which one is \btagged) when a  combination of the two photons and another jet passes the Top1 condition (see text).
To normalise the simulation to data, the \Syyj\ contribution is scaled up by 3\% and down by 4\% in (a) and (b), respectively.
The signal corresponds to the $tcH$ coupling, with a $t\to cH$ branching ratio of 2\%. The hatched bands represent the statistical uncertainty in the simulated background. The vertical dotted lines indicate the ranges of the Top1 and Top2 invariant mass selections.}
\end{figure}
 
The had-tH category, which targets the $pp \to tH$ production, is populated with events with three jets and events with four or more jets that are not retained in the above categories.
All jets are subject to \btag and at least one 3-jet combination meeting the Top2 mass constraint is required, with the additional requirements that one of the three jets is \btagged, and there is no additional \btagged jet.
 
To improve the sensitivity of the analysis, a multivariate selection is performed, using boosted decision trees (BDT) as implemented in TMVA~\cite{TMVA}.
 
\subsubsection{Additional BDT selection}
For the four categories addressing  the \nttb\ production, the BDT is trained with events from both sub-categories $c$ and $\cancel{c}$ together, as the response of the charm-tagging algorithm is largely independent of the kinematic properties of the objects entering the BDT\@. The BDT is trained using the $tcH$ signal and non-resonant background processes, as described in section~\ref{sec:MC}, with a diphoton invariant mass
limited to the range [115,~135]~\gev. This range is chosen to select backgrounds in the signal mass range, while keeping enough events for the BDT training.
 
Several combinations of jets can satisfy the selection requirements for the same analysis category.
The combination for which \mggj\ is closest to 171~\gev, and \mjjj\ is closest to 170~\gev\
is used to build the BDT input variables.
Starting from a large number of BDT input variables, a reduced working set is obtained by removing the least discriminating variable until a marked decrease of the significance~\footnote{Here, the significance is defined as $\sqrt{2(s+b)\ln(1+s/b)-2s}$, where $s$ is the FCNC signal yield, assuming ${\cal B} = 10^{-3}$, and $b$ the non-resonant background yield in the diphoton mass range from 122~\gev\ to 129~\gev.} is observed.
For both the \nttb- and $tH$-targeted categories a set of seven variables was chosen.
For the categories targeting the \nttb\ channel, the variables used as input to the BDT, ranked in decreasing order of sensitivity, are:
\begin{enumerate}
\item $\pt^{\gamma\gamma}$ : transverse momentum of the diphoton system;
\item $m_{jj}$ : invariant mass of the $W$-boson candidate, defined as the two-jet system, that,
combined with the \bjet candidate, forms the $t \to Wb$ decay candidate;
\item $m_{tt}$ : invariant mass of the two top-quark candidates;
\item $\pt^{b}$ : transverse  momentum of the \bjet candidate;
\item max$(\Delta R_{c\gamma})$ : distance between the jet candidate in Top1 and the farthest photon;
\item min$(\Delta R_{bj})$ : distance between the \bjet candidate and the closest jet;
\item $\HT^{\gamma\gamma 4j}$ : scalar sum of the \pt of the two photons and the four jets of the retained combination.
\end{enumerate}
The distributions of $\pt^{\gamma\gamma}, m_{tt}, \pt^{b}$ and $\HT^{\gamma\gamma 4j}$ extend to larger values for the signal than for the dominant \Syyj\ background. For signal events max$(\Delta R_{c\gamma})$ and min$(\Delta R_{bj})$ are on average smaller than for background, and $m_{jj}$ peaks near \mW. The variables $m_{jj}$ and $m_{tt}$ are not used for the had-tx categories.
 
For the tH category, the BDT is trained using the $tuH$ signal sample, which populates the tH-selected sample in similar proportions
from the $tH$ and \ttbar production modes, as opposed to the $tcH$ signal, where the contribution from the former is much reduced,
due to a smaller $pp \to tH$ production cross-section from a charm quark. The background events
used for the training are the same as the ones used for the \ttbar-targeted BDTs.
The selected variables, ranked in decreasing order of sensitivity, are:
\begin{enumerate}
\setcounter{enumi}{0}
\item $\pt^{\gamma\gamma}$;
\item $m_{jj}$;
\item $\HT^{\mathrm{Top2}}$: scalar sum of the \pt of the three jets entering the Top2 combination;
\item $\HT^{\mathrm{jets}}$ : scalar sum of the \pt of all selected jets;
\item min$(\Delta R_{bj})$;
\item $(\pt^{\gamma_1}+\pt^{\gamma_2})/\mgg$: scalar sum of the \pt of the two photons normalised to the diphoton invariant mass;
\item $\Delta R_{bW}$ : distance between the \bjet candidate and the $W$-boson candidate.
\end{enumerate}
Three of the seven variables are the same as in the \nttb\ case. Similarly to min$(\Delta R_{bj})$, $\Delta R_{bW}$ tends to have lower values for signal than for background. The distributions of the variables $\HT^{\mathrm{Top2}}, \HT^{\mathrm{jets}}$ and $(\pt^{\gamma_1}+\pt^{\gamma_2})/\mgg$ are slightly harder for signal than for background. The normalisation to \mgg\ for the latter variable prevents the BDT from learning that the signal is narrowly peaked in \mgg. For events with three selected jets, $\HT^{\mathrm{Top2}}$ and $\HT^{\mathrm{jets}}$ are identical.
 
Events with a BDT score (between $-1$ and 1) larger than a given threshold are retained. This threshold is determined
by maximising the expected significance computed using only events satisfying ${122\ \gev < \mgg < 129\ \gev}$.
For the tH category two thresholds are used to optimise the performance for both the $tuH$ and the
$tcH$ signals. Events with a BDT score
larger than 0.45 enter the tHT category, most relevant for $tuH$, while events with
a score between 0.2 and 0.45 enter the tHL category, more relevant for $tcH$.
The significance improvements are about 20\% for the \ttbar-targeted categories, and 30\% for the $tH$-targeted ones.
The signal acceptances, with sub-categories $c$ and $\cancel{c}$ grouped, are shown in table~\ref{tab:accep-had}.
\begin{table}[h!]
\begin{center}
\caption{Acceptance, in percent, of the hadronic selection for simulated signal events. The sub-categories $c$ and $\cancel{c}$ are grouped. The acceptances for the tH category with BDT selection are given for tHL and tHT together. The uncertainties are statistical only.}
\label{tab:accep-had}
\small
\begin{tabular}{l|ccc|ccc}
\hline
Selection    &  \multicolumn{3}{c|}{Before BDT selection} & \multicolumn{3}{c}{After BDT selection} \\
Category   &   had-tt &   had-tx  &  had-tH   & had-tt & had-tx & had-(tHL+tHT) \\
\hline
$tcH$ & 2.81$\pm$0.02 & 2.08$\pm$0.01 & 3.50$\pm$0.02 & 2.13$\pm$0.01 & 1.14$\pm$0.01 &1.80$\pm$0.01 \\
$tuH$ & 2.23$\pm$0.01 & 1.66$\pm$0.01 & 3.55$\pm$0.02 & 1.58$\pm$0.01 & 0.94 $\pm$0.01 & 2.01$\pm$0.02 \\
\hline
\end{tabular}
\end{center}
\end{table}
 
\subsubsection{Diphoton invariant mass distributions}
The distributions of the diphoton invariant mass, for each of the six
categories, after the BDT selection, are shown in figure~\ref{hadronic:fig:myy}.
The contribution of the $tX\gamma$ and $V\gamma\gamma$ processes
are normalised to the cross-sections as predicted by the MC generators and the integrated luminosity of the sample.
The \Syyj\ normalisation is obtained by scaling this sample, in such a way that the total number of simulated events outside of the [120, 130] \gev\ \mgg\ interval matches the number observed in data, for each channel independently. The FCNC signal contributions correspond to ${\cal B} = 5 \times 10^{-4}$.
\begin{figure}[htbp!]
\begin{center}
\hspace{-0.2 cm}
\subfigure[]{\includegraphics[width=.45\textwidth]{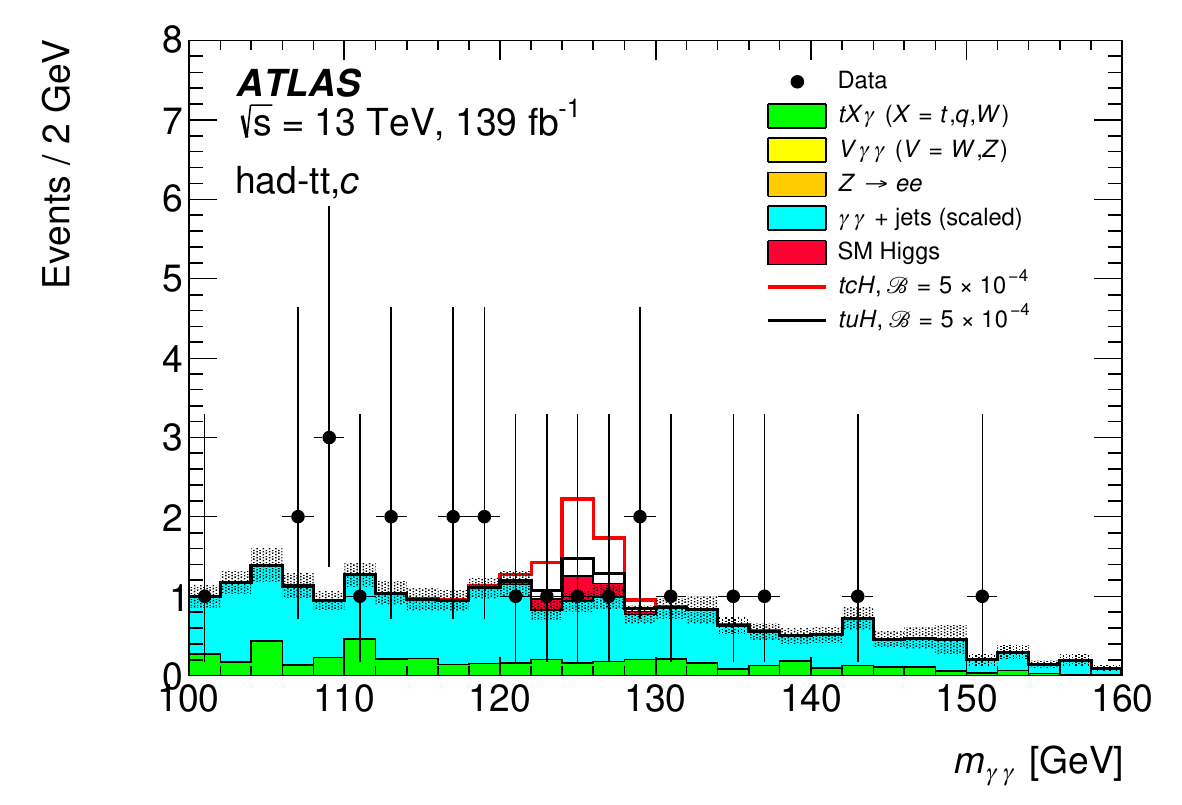}}
\subfigure[]{\includegraphics[width=.45\textwidth]{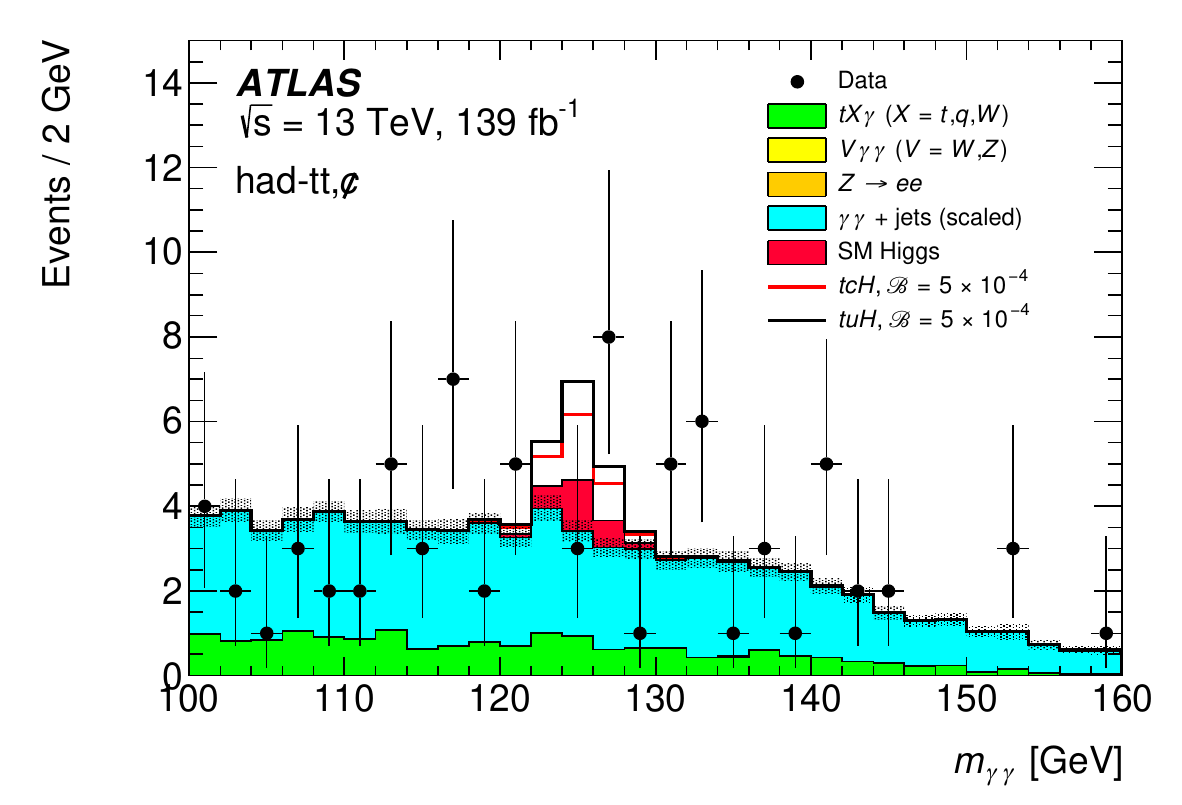}}
\subfigure[]{\includegraphics[width=.45\textwidth]{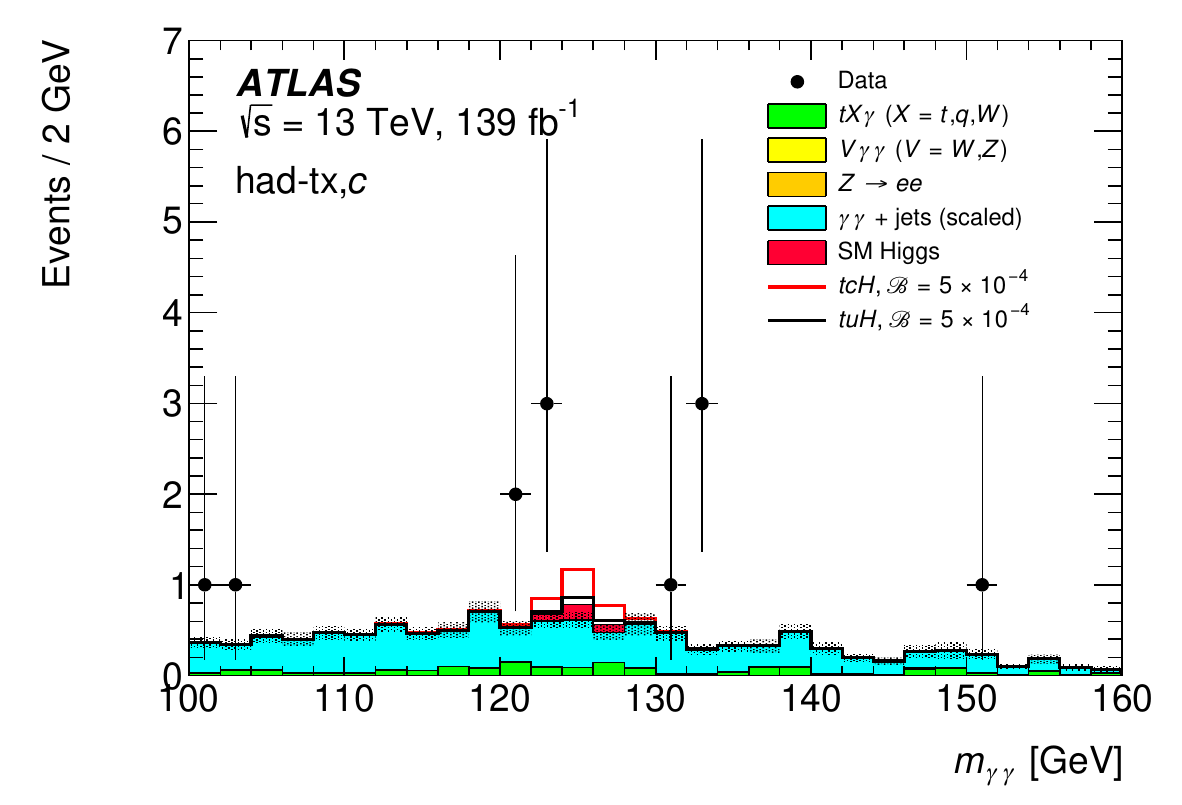}}
\subfigure[]{\includegraphics[width=.45\textwidth]{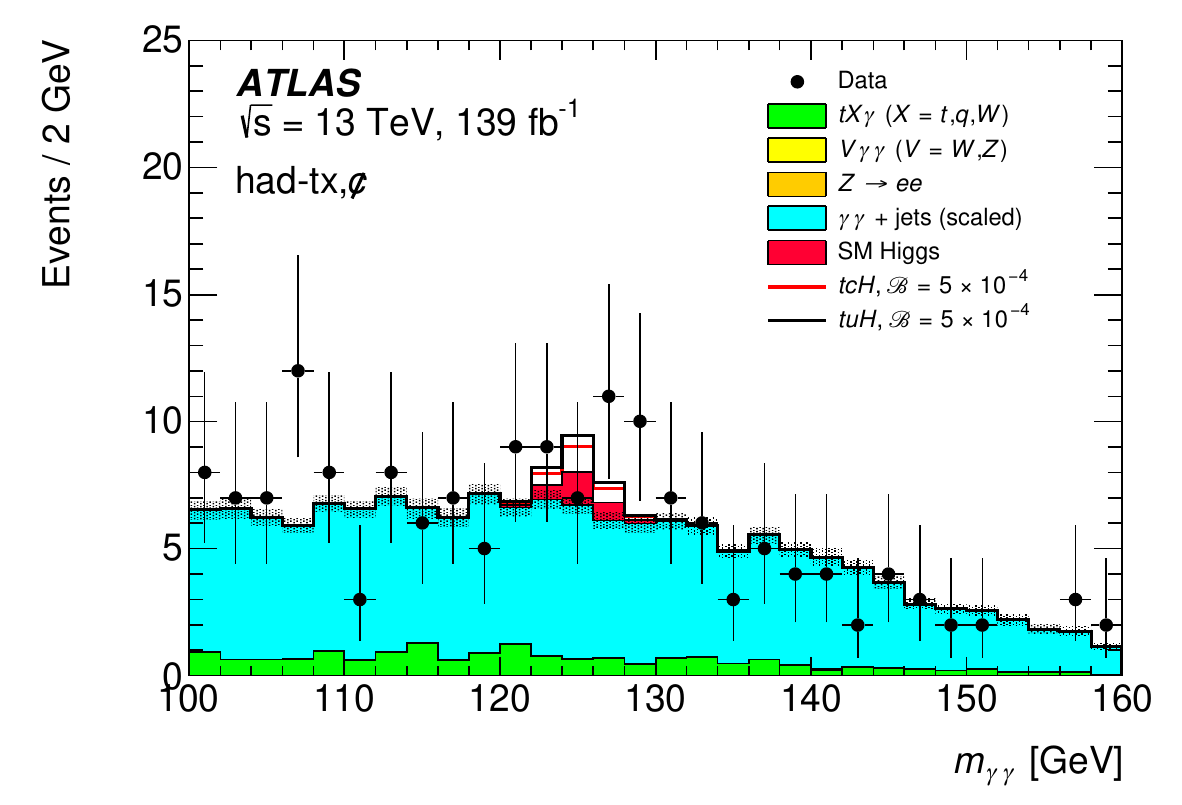}}
\subfigure[]{\includegraphics[width=.45\textwidth]{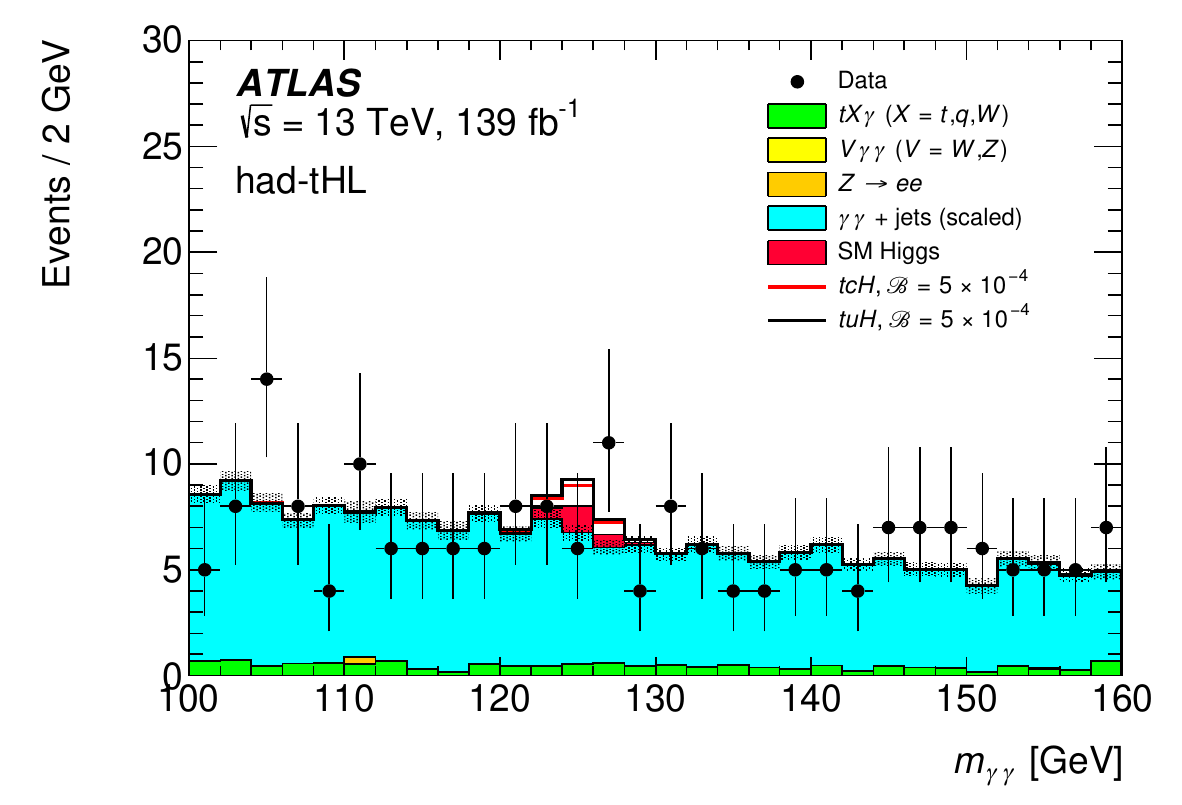}}
\subfigure[]{\includegraphics[width=.45\textwidth]{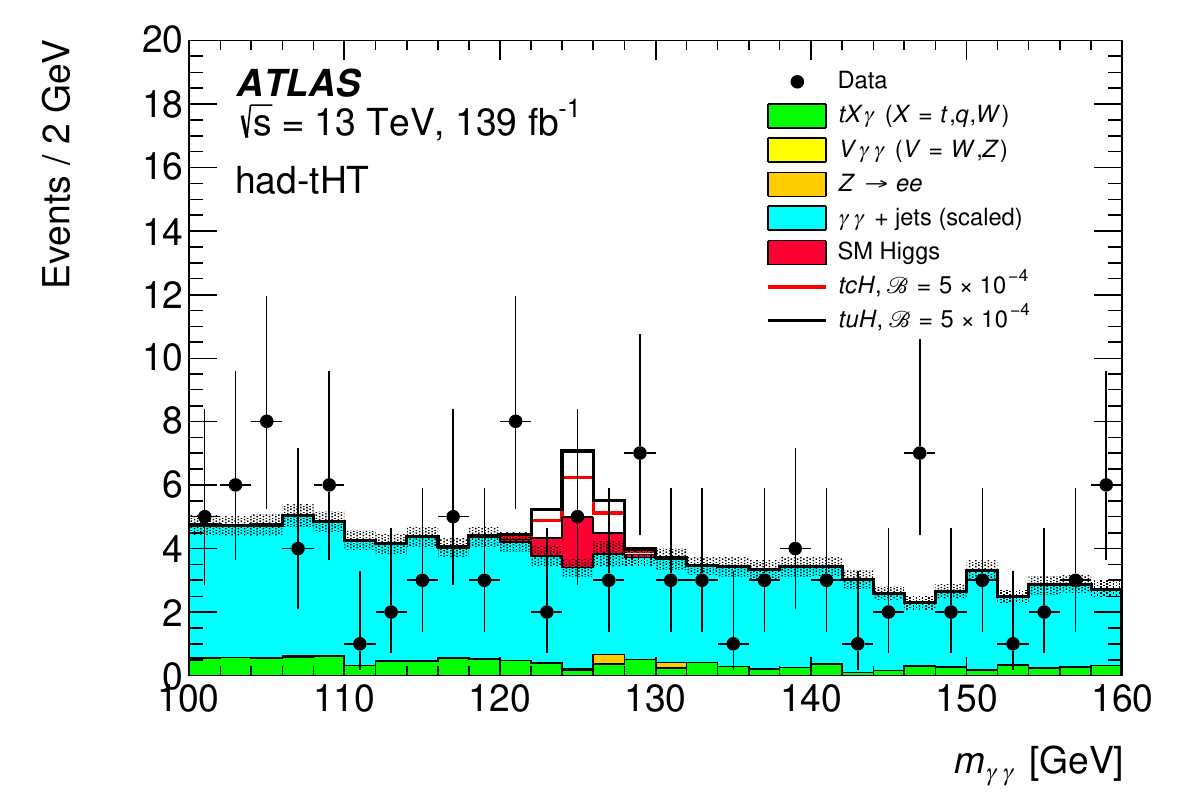}}
\end{center}
\vspace{-0.5cm}
\caption{Distributions of the diphoton invariant mass for data, signal, Higgs boson production in the SM, and non-resonant background for the (a) had-\uA, (b) had-\uB, (c) had-\dA, (d) had-\dB\ categories after the BDT selection, and (e) had-tHL and (f) had-tHT categories. The hatched bands correspond to the statistical uncertainty in the sum of the simulated non-resonant backgrounds.}
\label{hadronic:fig:myy}
\end{figure}
 
\subsection{Leptonic selection}
\label{subsec:Leptonic}
The leptonic selection targets the processes $\nttb \to bW(\ell\nu)\bar{q}H(\gamma\gamma)$ and $tH\to bW(\ell\nu)H(\gamma\gamma)$.
Exactly one lepton and one or more jets are required, in addition to the two photons.
Figure~\ref{selection:fig:mtrans} shows the transverse mass, \mt,
of the $W$-boson candidate, calculated from the transverse momentum of the lepton and the missing transverse momentum.
An additional requirement of exactly one \btagged jet is applied.
The sum of the non-resonant backgrounds ($tX\gamma, V\gamma\gamma$ and \Syyj) is shown together with data.
A $Zee$ component, resulting from $Z \to ee$ events, in which one of the electrons is misidentified as a photon candidate and the other photon is genuine, is also considered.
Its normalisation is fixed using the electron-photon (leading and sub-leading) mass distributions, which show an enhancement at the $Z$-boson mass. Altogether, the different contributions provide a relatively good description of the background although in the low-\mt\ region, the simulation overestimates the data.
\begin{figure}[h!]
\begin{center}
\includegraphics[width=.5\textwidth]{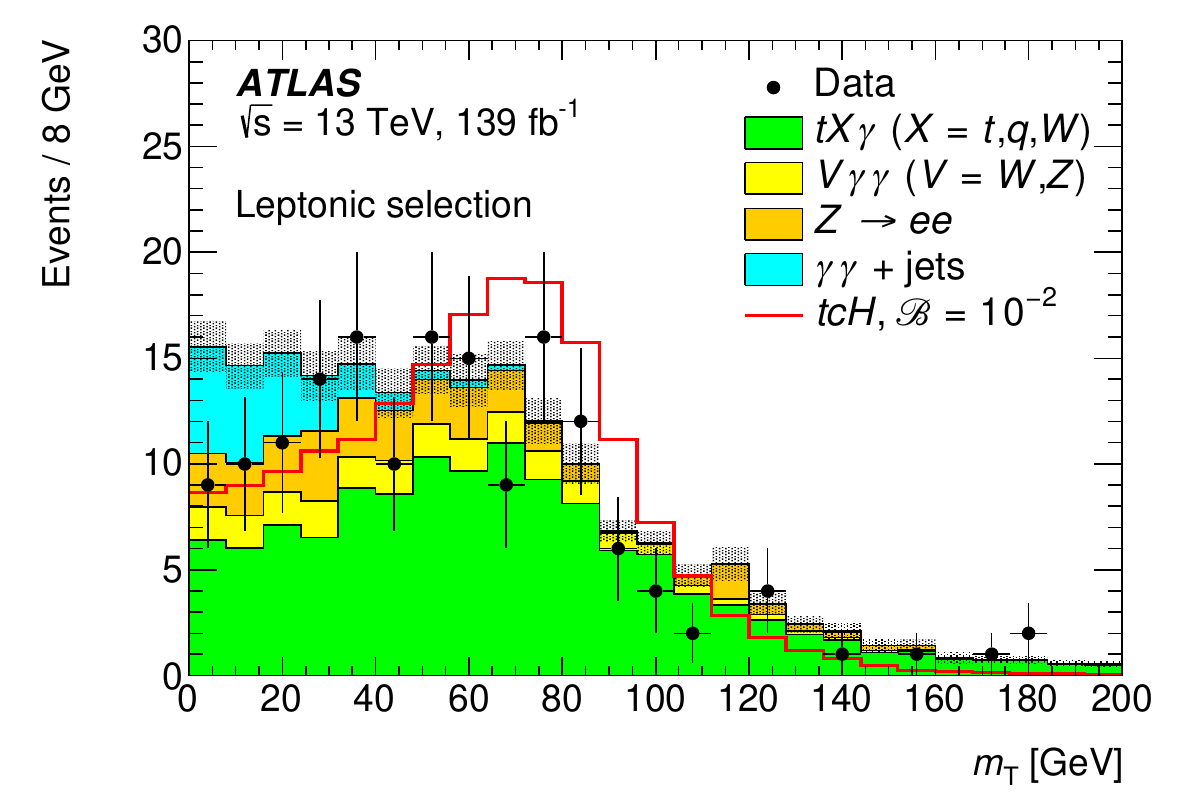}
\end{center}
\vspace{-0.5cm}
\caption{Distribution of the transverse mass of the lepton and \met, for events
with exactly one charged lepton and one \btagged jet. The distributions of the main backgrounds and of a $tcH$ signal
with a branching ratio of 1\% are also shown.
The statistical uncertainty in the sum of the non-resonant backgrounds is represented as a hatched band.}
\label{selection:fig:mtrans}
\end{figure}
 
Using a $W$-boson mass constraint allows for the calculation of the longitudinal momentum of the escaping neutrino, and therefore the reconstruction
of the invariant mass \mlnj\ of the $\ell\nu j$ system (Top2).
To ensure a reliable reconstruction of this mass and to reject background,
the events are required to satisfy $\mt > 30~\gev$,
except for those collected in the very loose category lep-R, defined later.
If there are two valid solutions for the longitudinal momentum of the neutrino, the solution giving a Top2 mass closer to 170 \gev\ is retained. In the absence of a real solution, \mW is replaced by ${\mt + 100\ \mev}$ in the mass constraint, which ensures two almost degenerate, real solutions.
 
\subsubsection{Categories}
Addressing first the $\nttb$ production, the \mggj\ invariant mass distribution is formed, considering each of the up to five jets in the event.
The distribution is shown in figure~\ref{selection:fig:mtop12_lepa}. All jets that satisfy $152\ \gev < \mggj\ < 190\ \gev$ are subject to
\ctag (and not to \btag). All other jets are subject to \btag. There must be exactly one \btagged jet among those eligible.
After the requirement that at least one $\gamma\gamma j$ combination passes the Top1 condition, the distribution of the \mlnj\ invariant mass is shown in figure~\ref{selection:fig:mtop12_lepb}. Only one entry is made if more than one jet could be associated with the two photons and satisfy the Top1 condition.
\begin{figure}[h!]
\begin{center}
\hspace{-0.2 cm}
\subfigure[]{\includegraphics[width=.5\textwidth]{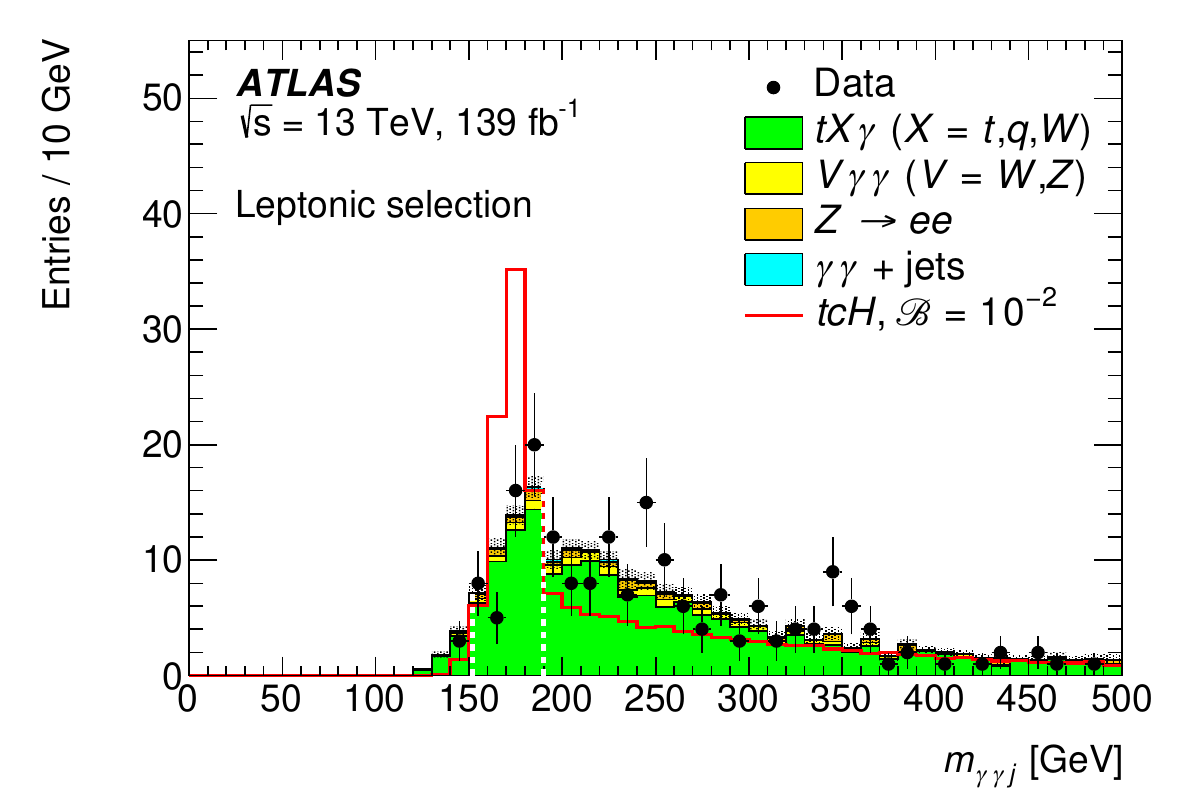}\label{selection:fig:mtop12_lepa}}
\subfigure[]{\includegraphics[width=.5\textwidth]{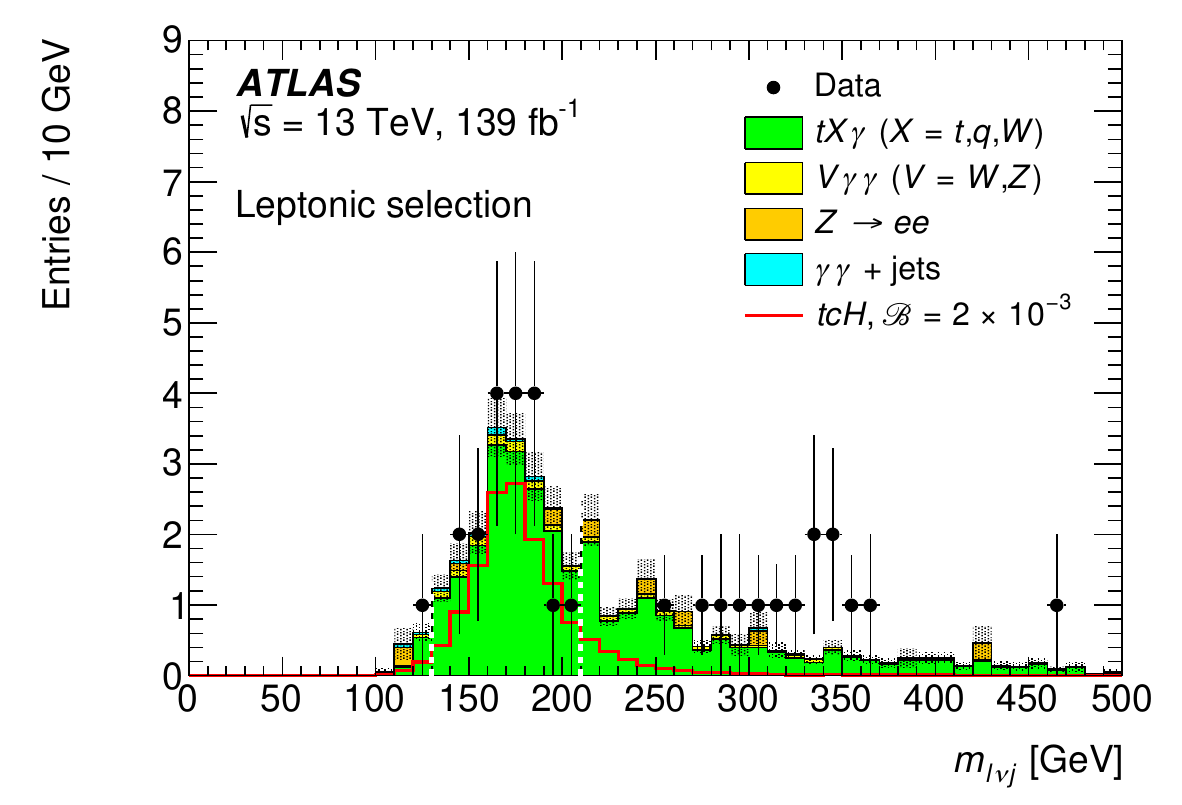}\label{selection:fig:mtop12_lepb}}
\end{center}
\vspace{-0.5cm}
\caption{Distributions of (a) the invariant mass of the two photons and one jet, with the additional condition that there is one \btagged jet among the other jets that can form a Top2 candidate with the lepton and \met, and (b) \mlnj\ ($j$ is the \btagged jet) when at least a combination of the two photons and another jet passes the Top1 condition. The signal corresponds to the $tcH$ coupling with a 1\%  or 0.2\% $t \to cH$ branching ratio for (a) and (b), respectively. The simulation (except for $Zee$) is rescaled to match the data by a factor of 0.96 and 0.83 in (a) and (b), respectively.
The statistical uncertainty in the sum of the non-resonant backgrounds is represented as a hatched band.
}
\end{figure}
 
Events with at least one combination fulfilling the Top1 condition, for which, in addition, the Top2 requirement $130\ \gev < \mlnj < 210\ \gev$ is met
with the \btagged jet, are assigned to category lep-tt. If the jet in the $\gamma\gamma j$ combination meeting the Top1 condition is \ctagged, the event is assigned to the lep-\uA\ category and otherwise to the lep-\uB\ category.
 
The lep-tH category, which targets the $pp \to tH$ production, contains events with only one jet, and events with two or more jets for which no Top1 condition is satisfied. All jets are subject to \btag, and there must be exactly one \btagged jet among the (up to five) leading ones. The \mlnj\ invariant mass is required to satisfy the Top2 condition, with the \btagged jet. In the lep-tH category, the background is still reducible, and a further selection is made using a BDT\@.
Similarly to the had-tH categories, the signal originating from the $tuH$ coupling is used for training. The background sample is the sum of the $tX\gamma$, $V\gamma\gamma$ and \Syyj\ contributions. To have enough events in the background sample, the full diphoton mass range from 100~\gev\ to 160~\gev\ is used for the training.
The seven most sensitive variables used as input to the BDT, ranked in decreasing order of sensitivity, are:
\begin{enumerate}
\item $\pt^{\gamma\gamma}$;
\item $m_{\mathrm{T}}$ : the transverse mass of the lepton+\met system;
\item $Q_{\mathrm{lep}}$ : the lepton's electric charge;
\item N$_{\mathrm{jets}}$ : the number of jets passing the \pt and $y$ requirements;
\item \HT : the sum of \met, lepton \pt and \pt of the all jets passing the jet selection;
\item $y_{\gamma\gamma}$ : the rapidity of the diphoton system;
\item $(\pt^{\gamma_1}+\pt^{\gamma_2})/\mgg$.
\end{enumerate}
Four of the seven retained variables are equivalent to the ones used in the hadronic tH selection. The value of N$_{\mathrm{jets}}$ is significantly smaller on average for the signal than for backgrounds, and $Q_{\mathrm{lep}}$ has twice as many positives than negatives, while $y_{\gamma\gamma}$ extends towards larger values, altogether providing additional separation power. For a $tcH$-originating signal, a BDT trained with a $tcH$ signal does not yield a better significance than the one trained with the signal originating from $tuH$, thus only the latter is used, with a single optimum lower threshold of 0.15. The BDT brings a significance improvement
of about 25\% (20\%) for a signal originating from $tuH$ ($tcH$).
 
All other events that have one lepton, two photons, and one \btagged jet --- excluding those categorised as lep-\uA, lep-\uB, or lep-tH after BDT selection --- are assigned to an additional category
named lep-R.

The signal acceptance is shown in table~\ref{tab:accep-lep}. Here, the categories $c$ and $\cancel{c}$ are grouped, as their sum is independent of the \ctag working point. Altogether, the total acceptance of the hadronic plus leptonic selections, adding the $\nttb$ and $tH$ production modes, is 8.3\% (7.4\%) for $tcH$ ($tuH$).
\begin{table}[h!]
\begin{center}
\caption{Acceptance, in percent, of the leptonic selection for simulated signal events. The sub-categories \uA\ and \uB\ are grouped. The uncertainties are statistical only.}
\label{tab:accep-lep}
\small
\begin{tabular}{l|ccc|c}
\hline
Selection          &  \multicolumn{3}{c|}{Before BDT selection} & \multicolumn{1}{c}{After BDT selection} \\
Category         &  lep-R  &   lep-tt  &  lep-tH   & lep-tH \\
\hline
$tcH$ & 1.42$\pm$0.01 & 1.09$\pm$0.01 & 1.11$\pm$0.01 & 0.69$\pm$0.01 \\
$tuH$ & 1.27$\pm$0.01 & 0.87$\pm$0.01 & 1.15$\pm$0.01 & 0.77$\pm$0.01 \\
\hline
\end{tabular}
\end{center}
\end{table}
\subsubsection{Diphoton invariant mass distributions}
The distributions of the diphoton invariant mass are shown in figure~\ref{leptonic:fig:myy}. The non-resonant background is normalised using the cross-sections as predicted by the MC generators, and the integrated luminosity. For all categories this gives an expected contribution in agreement with  data within statistical uncertainty. The FCNC signal contributions are normalised to ${\cal B} = 5 \times 10^{-4}$.
 
\begin{figure}[h!]
\begin{center}
\subfigure[]{\includegraphics[width=.38\textwidth]{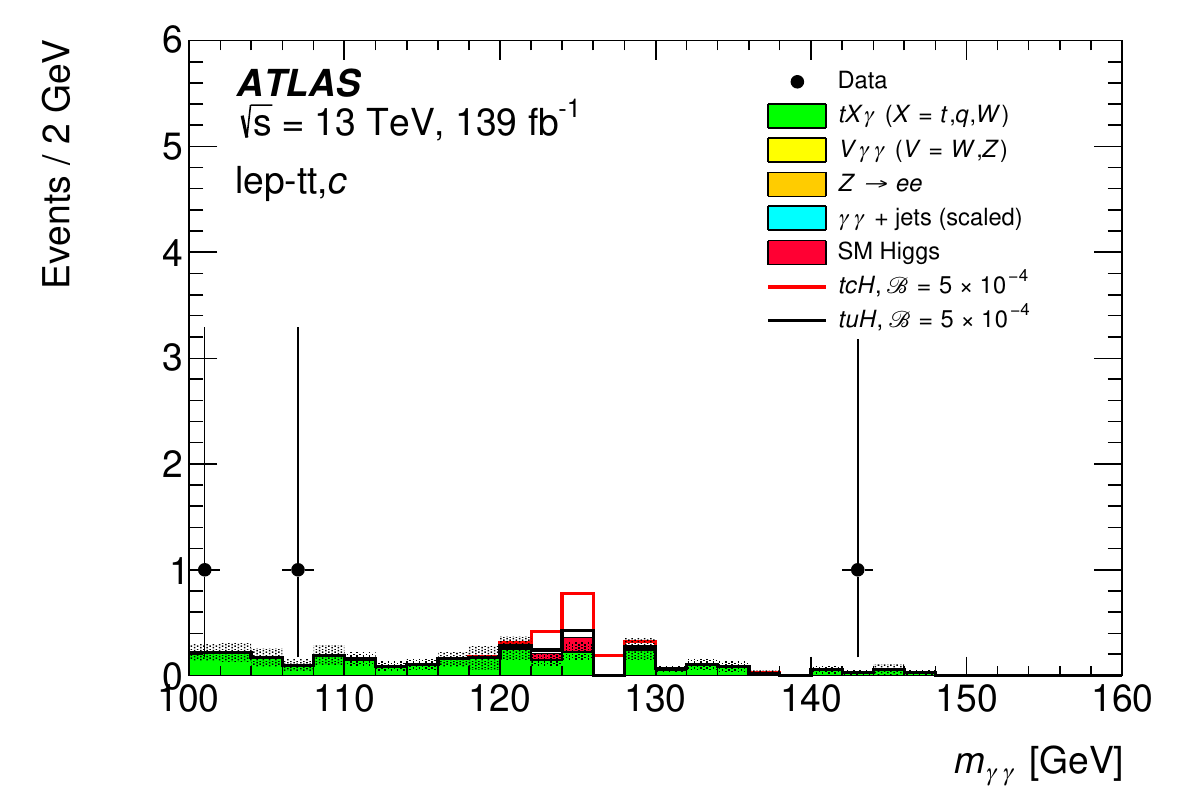}}
\subfigure[]{\includegraphics[width=.38\textwidth]{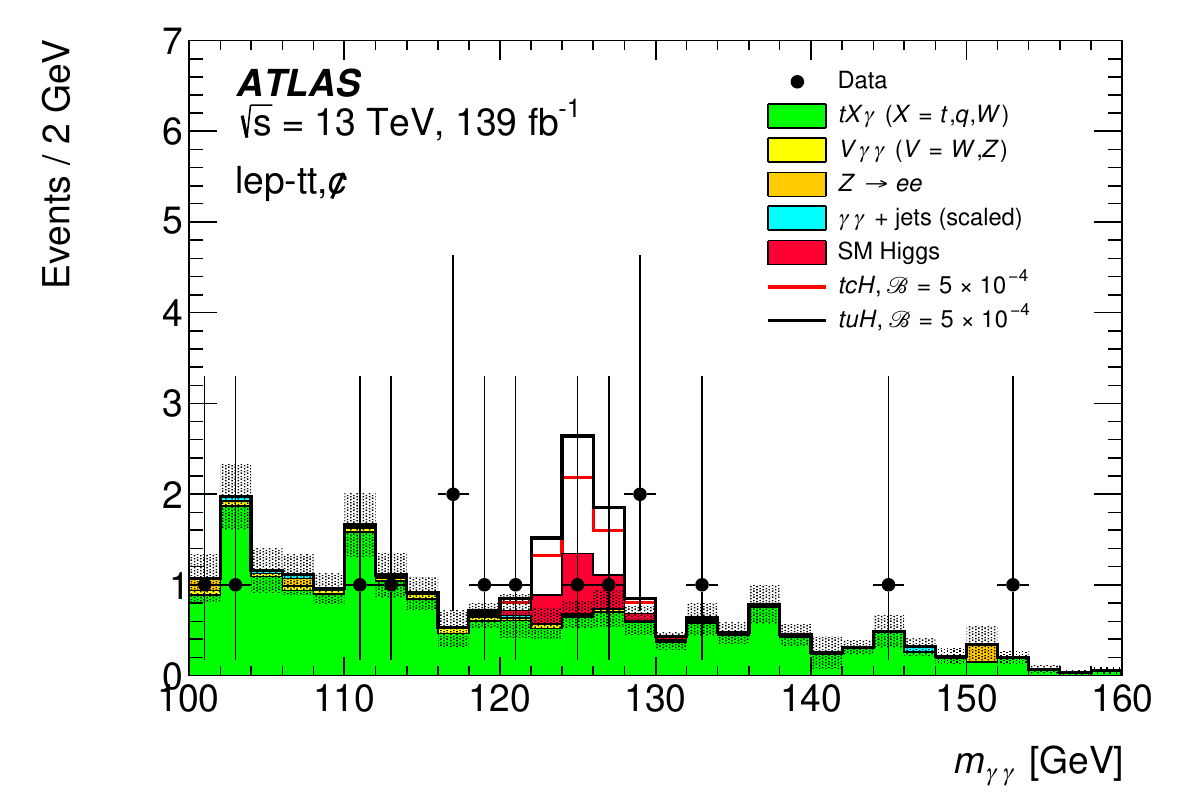}}
\subfigure[]{\includegraphics[width=.38\textwidth]{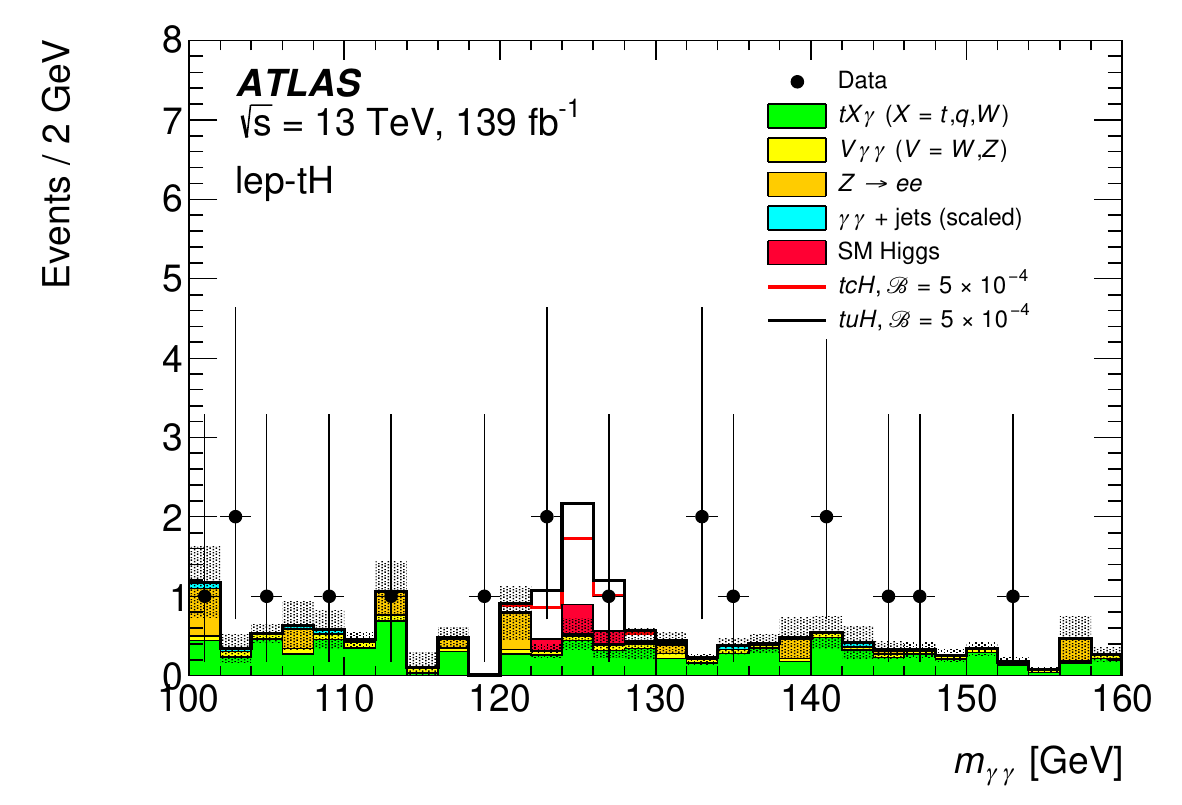}}
\subfigure[]{\includegraphics[width=.38\textwidth]{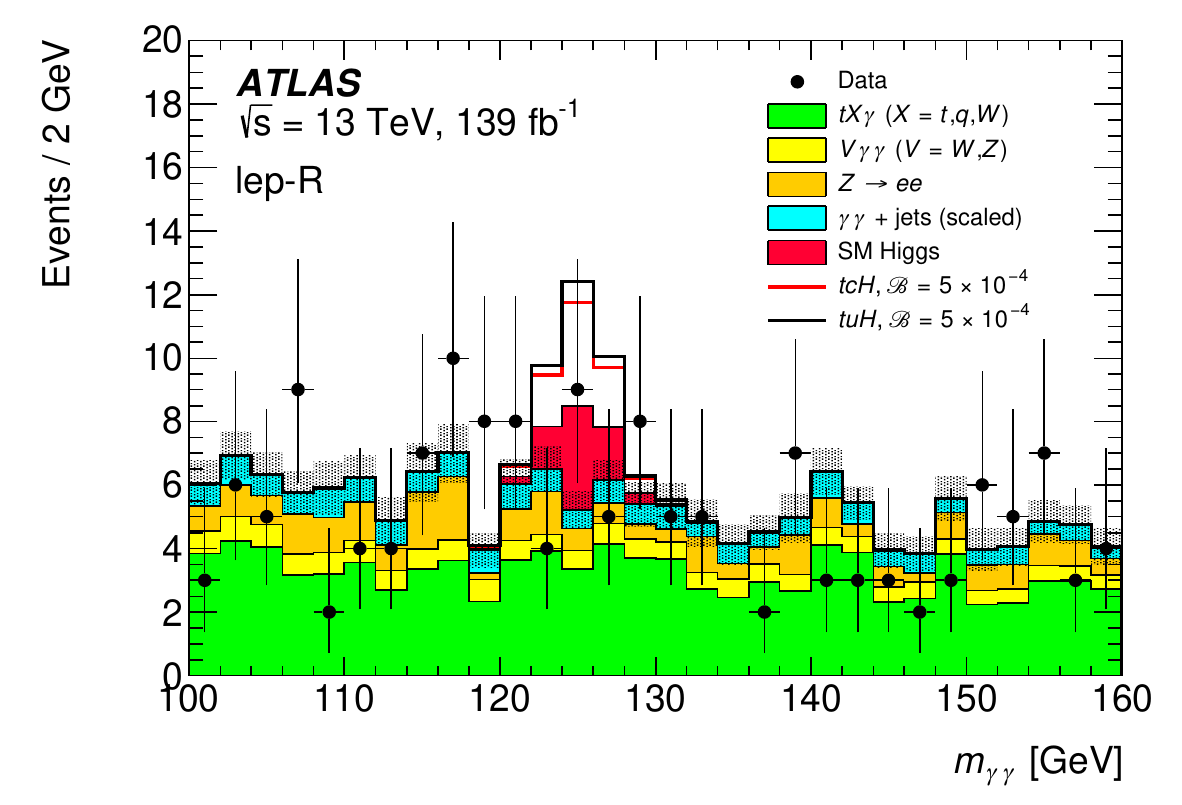}}
\end{center}
\vspace{-0.5cm}
\caption{Distributions of the diphoton invariant mass for data, signal, Higgs boson production in the SM, and non-resonant background for events in categories (a) lep-\uA, (b) lep-\uB, (c) lep-tH after the BDT selection and (d) lep-R. The hatched bands correspond to the statistical uncertainty in the sum of the simulated non-resonant contributions.}
\label{leptonic:fig:myy}
\end{figure}
\FloatBarrier
\subsection{Summary of the selections and expected number of signal events}
\label{subsec:summary}
Figure~\ref{fig:selsummary} summarises the selection criteria described in the previous sections,
and table~\ref{tab:signal} shows the numbers of events expected in the ten analysis categories.
\begin{figure}[h!]
\begin{center}
\subfigure[]{\includegraphics[width=.49\textwidth]{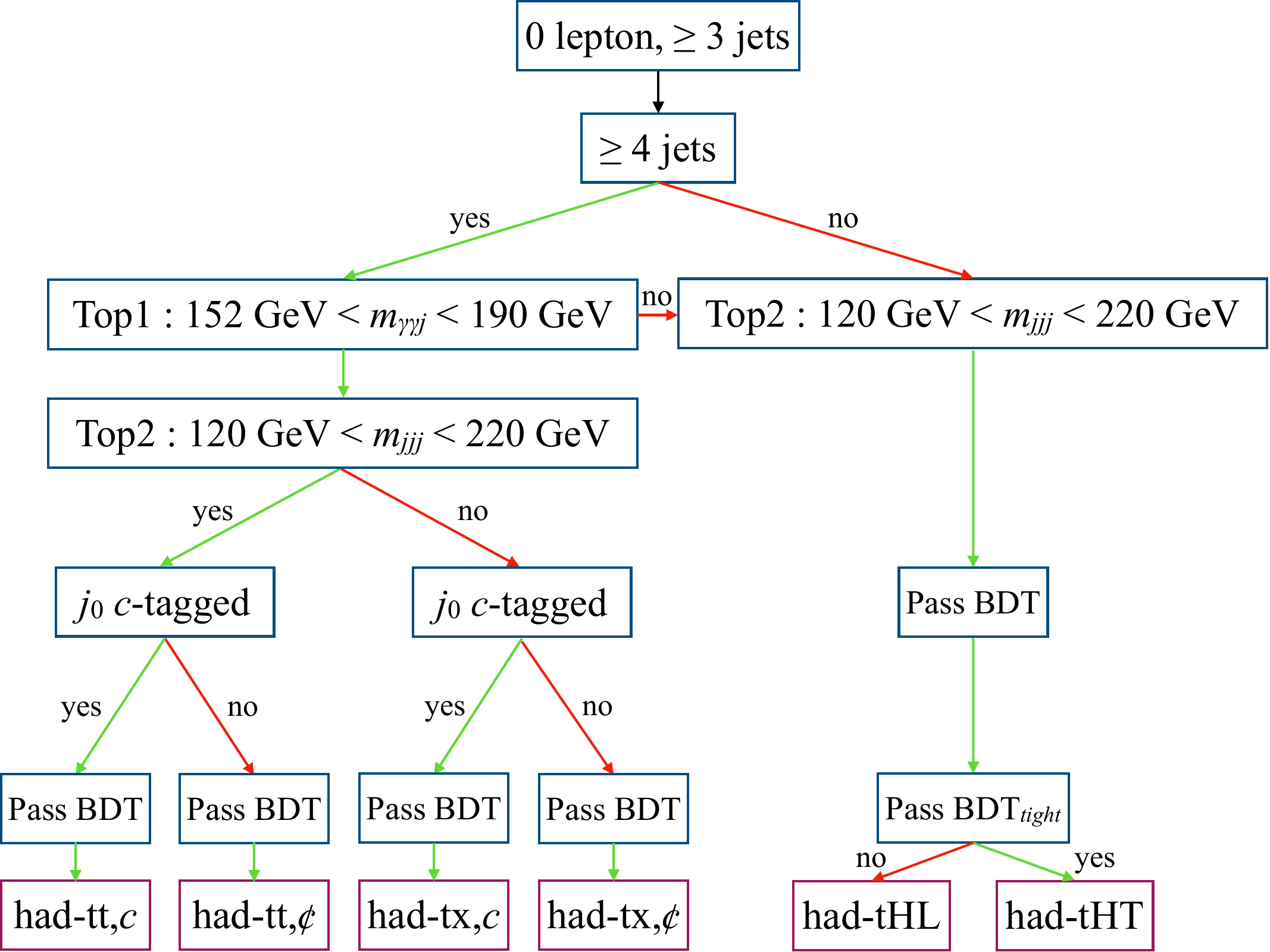}}
\subfigure[]{\includegraphics[width=.49\textwidth]{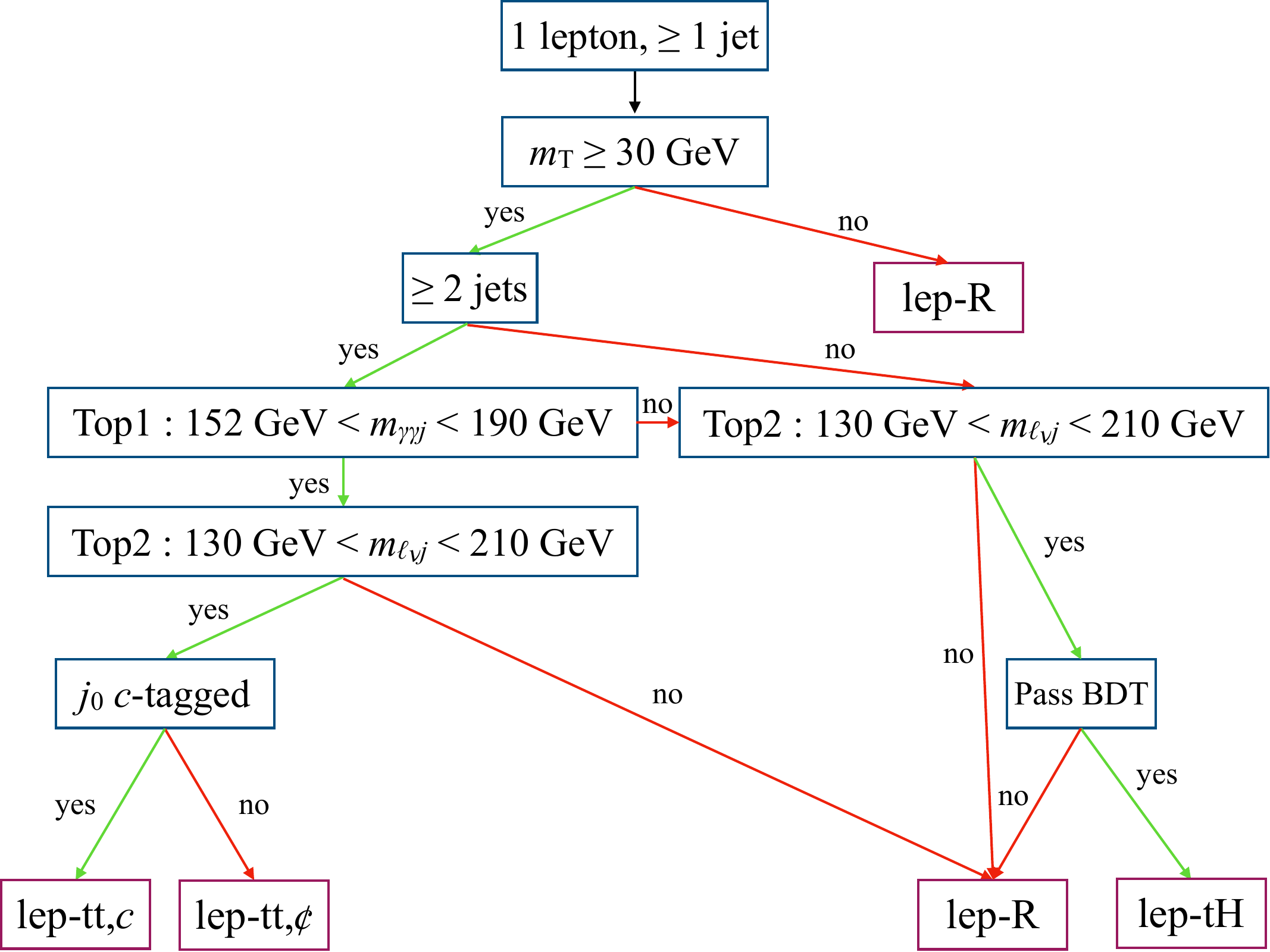}}
\end{center}
\vspace{-0.5cm}
\caption{Summary of the categories used in the statistical analysis for the (a) hadronic and (b) leptonic channels.
Green (red) arrows indicate that the criterion in the box is met (not met).
Exactly one \bjet is required. The jet $j_0$, combined with the diphoton system, forms the $t \to qH$ decay candidate.}
\label{fig:selsummary}
\end{figure}
 
\begin{table}[h!]
\begin{center}
\caption{Number of signal events expected after the BDT selection. A branching ratio of $5 \times 10^{-4}$, close to the 95\%~CL expected limit in the absence of signal,
is assumed for the FCNC decay. The categories had-tHL and had-tHT are grouped in this table.
The uncertainties include the statistical uncertainty from the number of selected events, the
theoretical uncertainties in the $pp\to\nttb$ and the $pp\to tH$ cross-sections and in the $H \to \gamma\gamma$ branching ratio.}
\label{tab:signal}
\footnotesize
\begin{tabular}{p{11mm}|>{\centering}p{11mm}>{\centering}p{11mm}>{\centering}p{11mm}>{\centering}p{11mm}>{\centering}p{19mm}|>{\centering}p{11mm}>{\centering}p{11mm}>{\centering}p{11mm}>{\centering\arraybackslash}p{11mm}}
\hline
Selection &  \multicolumn{5}{c|}{Hadronic}      & \multicolumn{4}{c}{Leptonic} \\
Category  &  had-\uA  &  had-\uB    &  had-\dA  &  had-\dB   & had-\scriptsize{(tHL+tHT)}   &  lep-\uA   &  lep-\uB  & lep-tH     & lep-R\\  \hline
$tcH$     & 2.32$\pm$0.15 & 3.62$\pm$0.24 & 0.90$\pm$0.06 & 2.30$\pm$0.15 & 5.04$\pm$0.30 & 1.00$\pm$0.07 & 2.05$\pm$0.14 & 1.92$\pm$0.11 & 3.94$\pm$0.24\\
$tuH$    & 0.54$\pm$0.04  & 5.39$\pm$0.34 & 0.18$\pm$0.01 & 3.35$\pm$0.21 & 7.56$\pm$0.45 & 0.17$\pm$0.01 & 3.08$\pm$0.20 & 2.90$\pm$0.15 & 4.74$\pm$0.26\\
\hline
\end{tabular}
\end{center}
\end{table}

% End of text imported from the .//Sections/EvtRecoSel.tex input file

\section{Statistical analysis and systematic uncertainties}

% The next lines are included from the .//Sections/StatSys.tex input file
The branching ratio $\cal B$ of the decay $t \to c(u)H$
is determined in a fit to data by using a likelihood function $L(\cal B)$, defined as the product of the likelihoods for
the ten categories. Hypothesised values of $\cal B$ are probed with a test statistic based on a
profile likelihood ratio. The theoretical uncertainties are mainly related to the \nttb\ and $tH$ production cross-sections,
the $H \to \gamma\gamma$  branching ratio, the resonant background from SM processes and
the signal generator uncertainties. The estimate of the uncertainty in the non-resonant background is
explained in section~\ref{sec:stat:liknonres}. The experimental and the event generator systematic uncertainties are
detailed in section~\ref{sec:stat:syst}. All systematic uncertainties are introduced as nuisance parameters in the likelihood.
 
\subsection{Likelihood construction}
\label{sec:stat:liknonres}
Two different likelihood constructions are used according to
the number of events observed in the side band (SB) of the \mgg\ mass distribution (defined as $\mgg \in [100,122[ \cup ]129,160]\ \gev$).
A too small number prevents a reliable fit of the background
shape parameters. This is the case for the lep-\uA\ category,
for which simple event counting is used. Here, the number of non-resonant background events
in the signal region (SR defined by $\mgg \in [122,129]\ \gev$) is estimated as $b^{SR} = \alpha
b$, where $b$ is the yield of the non-resonant background in the side-band region, and $\alpha$
is a transfer factor discussed below. For the nine other categories, the number of events in the SB is enough
to determine the background shape and a standard unbinned extended likelihood is used.
 
The resonant diphoton invariant mass distributions, assumed to be the same for the signal and SM Higgs boson processes,
are described by double-sided Crystal-Ball~\cite{Oreglia:1980cs} (DSCB) functions (a Gaussian function with power-law tails
on both sides), whose parameters are obtained from a fit to the mass spectrum of the simulated FCNC signal.
 
For all categories but lep-\uA, the background shapes are determined
using the \mgg\ distributions built from the simulation (MC templates). A set of smooth templates
is additionally built using
kernel smoothing~\cite{Cranmer:2000du} applied to the background distributions.
Fits of a signal + background model to the templates are performed using an analytic function for the background model.
The Higgs boson mass hypothesis is scanned in the range [121, 129]~\gev\ in steps of 0.5~\gev\ and a background functional form is accepted
if the condition $n_{sp} < 0.2\sigma_{bkg}$ is met for all \mH, where $n_{sp}$ is the fitted spurious signal and $\sigma_{bkg}$
is the statistical uncertainty in the fitted signal yield. The preference is given to the raw templates, for which a relaxed spurious
signal criterion (made to accommodate $2\sigma$ local fluctuations in the MC template, if the MC statistical uncertainty is large)
must be used in some cases, resulting in larger $n_{sp}$. However, the chosen functional form must also satisfy the standard criterion for
the smooth template. When several functions pass the spurious signal criteria, the one with the smallest number of parameters is chosen.
The functions chosen for the background model are a Fermi-Dirac function for category had-\uA, an exponential of second order polynomial for categories
had-\uB\ and had-tx, and a decreasing exponential function for the other categories; $n_{sp}$ ranges from 0.16 event for the had-\uA\ category
to 2.3 events for the lep-R category.
 
In the lep-\uA\ category, the transfer factor $\alpha$ is estimated as the average
obtained from two sources: a control data sample for which one of the two photons is either not isolated, or not tightly identified,
and the simulation sample, normalised to match the uncertainty from the data control sample. The resulting transfer factor is $\alpha= 0.27 \pm 0.15$,
where the uncertainty is dominated by the statistical uncertainty of the control sample.
For a uniform \mgg\ distribution $\alpha = 0.13$ is expected.
 
\subsection{Systematic uncertainties}
\label{sec:stat:syst}
 
\subsubsection{Theoretical uncertainties}
The uncertainties in the absolute normalisation of the signal yield take into account the uncertainties in the $pp \to \ttbar$ (6.1\%)
and $pp \to tH$ cross-sections. The latter were estimated with \MGNLO and the \mbox{TopFCNC} UFO at NLO, and amount
to 7.9\% (7.5\%) for the $tcH$ ($tuH$) coupling.
The uncertainty in the $H\to\gaga$ branching ratio~\footnote{The SM branching ratio, confirmed by the ATLAS and CMS collaborations
at the $\mathcal{O}(10)\%$ level~\cite{HIGG-2021-23,CMS-HIG-22-001}, is used.} (2.9\%~\cite{deFlorian:2016spz}) is included as well.
 
The uncertainties related to the choice of the factorisation and renormalisation scales in the signal acceptances are estimated
by varying the scales in the matrix element up and down by a factor of two relative to the  nominal values, and recalculating the
acceptance in each case. The largest deviation from the nominal acceptance is taken as the uncertainty.
The impact of the proton PDFs+$\alpha_s$ choice is obtained by using the root mean square of the signal acceptance
when considering the 100 MC replicas available in the \NNPDF[3.0nlo] set.
The uncertainty due to parton shower, underlying event and multiple interaction modelling is estimated
as the difference between the yields predicted from alternative signal samples, produced with \HERWIG~7.1.6~\cite{Bellm:2017jjp},
and those using the nominal samples produced with \PYTHIA~8.2.
Depending on the category, the corresponding uncertainties
range from 2\% to 16\% for the $pp \to \ttbar$ process and 1\% to 30\% for the $pp \to tH$ process; the largest uncertainties are
found in categories with a small acceptance for the considered process. This is the case, for example, for the latter number: it corresponds
to the variation of the yield of the process $pp \to tH \to \nqqbp bH$ in the lep-\uB\ category,
whose acceptance is below 2\% relative to the dominant contribution.
The impact of the \NLO tool choice for the generation of the
\ttbar events has been studied by comparing \POWPY~8.2 (nominal) and \MGNLOPY~8.2, for a subsample of events and the $tcH$ coupling.
The observed difference corresponds to an uncertainty ranging between 3\% and 10\%, depending on the category. The value of the
top-quark mass has a small impact on the acceptance of the tight criterion used to select Top1 candidates. The corresponding
uncertainty in the acceptance ranges from 0.3\% to 1.0\%, depending on the analysis category and on the production process.
 
SM cross-sections are assumed for the Higgs boson production.
An uncertainty in the global normalisation is taken into account (in addition to the
$H \to \gaga$ branching ratio uncertainty), which combines the QCD and PDF+$\alpha_s$ uncertainties as provided
in ref.~\cite{deFlorian:2016spz}.
 
\subsubsection{Experimental uncertainties}
The uncertainty in the integrated luminosity is 1.7\%~\cite{ATLAS-CONF-2019-021}, affecting the overall
normalisation of the signal and resonant background processes.
 
The systematic uncertainties in the photon energy scale and energy resolution have a large impact
on the signal shape. This impact is quantified from the relative variations of the mean
and interquartile range of the \mgg\ distribution between the nominal samples, and samples obtained by
varying the energy scale or energy resolution of the photon candidates according to their uncertainties.
They amount to approximately 0.5\% for the mean and up to 15\% for the width. The mean and width parameters
of the DSCB functions are allowed to float within these uncertainties. The impact
of the uncertainties in the photon energy scale and energy resolution on the signal yields is found to be negligible.
 
The uncertainties associated with the photon trigger and identification amount to about 2.8\%, dominated
by the identification efficiency uncertainty.
 
The differences in \btag and \ctag efficiencies between data and simulation are included in the event weights
of the simulated samples. The uncertainties in these weights induce variations of the expected signal yield
of the order of 1.5\% for \btag and up to about 10\% for \ctag in the \uA\ categories for a \ttbar signal with a
non-zero $tcH$ coupling. The uncertainties related to the JVT and fJVT selection are estimated in the same way.
 
The systematic uncertainties associated with the jet energy scale and jet energy resolution~\cite{JETM-2018-05}
are determined by modifying the jet
transverse momenta, re-running the analysis and estimating the yield variations in each category.
The changes in jet momenta are propagated to \met in this procedure. Adding in quadrature all effects,
the variations range from about 1\% to 9\%, depending on the category and the process (\ttbar or $tH$)
for categories where the signal yield is significant.
 
The uncertainty associated with the lepton energy scale, identification, and reconstruction efficiency
is smaller than 1\%. The uncertainty associated with \met is obtained with the same methodology
as that used for the jet energy scale, applied to the soft term introduced in section~\ref{subsec:otherobj}.
 
Since many of the uncertainties have a small impact and the analysis is statistically limited, uncertainties
affecting \btag, \met, photon identification and lepton-related reconstruction are grouped to
have only one effective nuisance parameter for each of these quantities.
 
The same uncertainties are included for the resonant background estimate. Since this background
mostly comes from the $pp \to \nttb H$ process, only this production mode is considered to assess
the size of the uncertainties, which are then used to scale the total yield from SM Higgs boson production.

% End of text imported from the .//Sections/StatSys.tex input file

\section{Results}

% The next lines are included from the .//Sections/Results.tex input file
\subsection{Constraints on the $tqH$ couplings using the $H \to  \gamma\gamma$ channel}
The diphoton mass spectra in the most significant categories are shown in figure~\ref{fig:myyShape},
together with the fitted background shapes and the resonant background and signal shapes, assuming a non-zero
$tcH$ coupling. No signal is observed, and the best fit estimate of the $t \to cH$ branching ratio, ${\hat{\cal B}}$, is
$(-1.3\pm2.3)\times 10^{-4}$, assuming the $tuH$ coupling to be zero. A summary of the fitted numbers of background
events, together with the numbers of observed events in the SR, is given in table~\ref{tab:summary}.
The yields for the standard Higgs boson production modes are constrained to the SM prediction within theoretical
and experimental uncertainties. The fitted $t \to uH$ branching ratio assuming the $tcH$ coupling to be zero is
$(-0.5\pm1.9)\times 10^{-4}$.
\begin{table}[h!]
\begin{center}
\caption{Numbers of events in the SR ($m_{\gamma\gamma} \in [122,129]\ \gev$) for the SM Higgs boson production
and the fitted non-resonant background, assuming a non-zero $tcH$ coupling, together with the number of
observed events in data, in the ten categories. The uncertainty in the total background yield in each category
is also shown.}
\footnotesize
\label{tab:summary}
\begin{tabular}{l|cccccc}
\hline
Selection     &  \multicolumn{6}{c}{Hadronic}  \\
Category                 & had-\uA   & had-\uB                 & had-\dA         &   had-\dB       & had-tHL           &  had-tHT        \\ \hline
SM Higgs boson      & \,\, 0.65  & \,\,\, 2.40               & \,\, 0.33         & \,\,\, 2.62      &   \,\,\, 2.37      &  \,\,\, 2.77    \\
Other background   &  3.9         & 12.4                         & 2.0                 & 24.3                &  23.0                & 13.2             \\ \hline
Total background    &  $4.5\pm1.0$ & $14.8\pm1.6$ & $2.3\pm0.6$ & $26.9\pm2.7$ & $25.4\pm1.7$ & $16.0\pm1.3$ \\ \hline
Data                        &  5 \,\,      & 11  \,\,                    & 3   \,\,           & 33     \,\,         &  27   \,\,           &  14 \,\,   \\ \hline
\end{tabular}
\\[0.2cm]
\begin{tabular}{l|cccc}
\hline
Selection     &   \multicolumn{4}{c}{Leptonic} \\
Category                 &  lep-\uA         & lep-\uB          & lep-tH          & lep-R               \\ \hline
SM Higgs boson      & \,\, 0.27         &   \,\,1.37       &   \,\, 0.68      &   \,\, 6.4          \\
Other background   &   0.5               &   1.6               &   2.2              & 17.2                 \\ \hline
Total background    & $0.8\pm0.6$ & $3.0\pm0.5$ & $2.8\pm0.5$ & $23.6\pm1.6$ \\ \hline
Data                        &   0  \,\,           &   3  \,\,          &   3  \,\,          &   22  \,\,          \\
\hline
\end{tabular}
\end{center}
\normalsize
\end{table}
 
\begin{figure}[h!]
\begin{center}
\subfigure[]{\includegraphics[width=.45\textwidth]{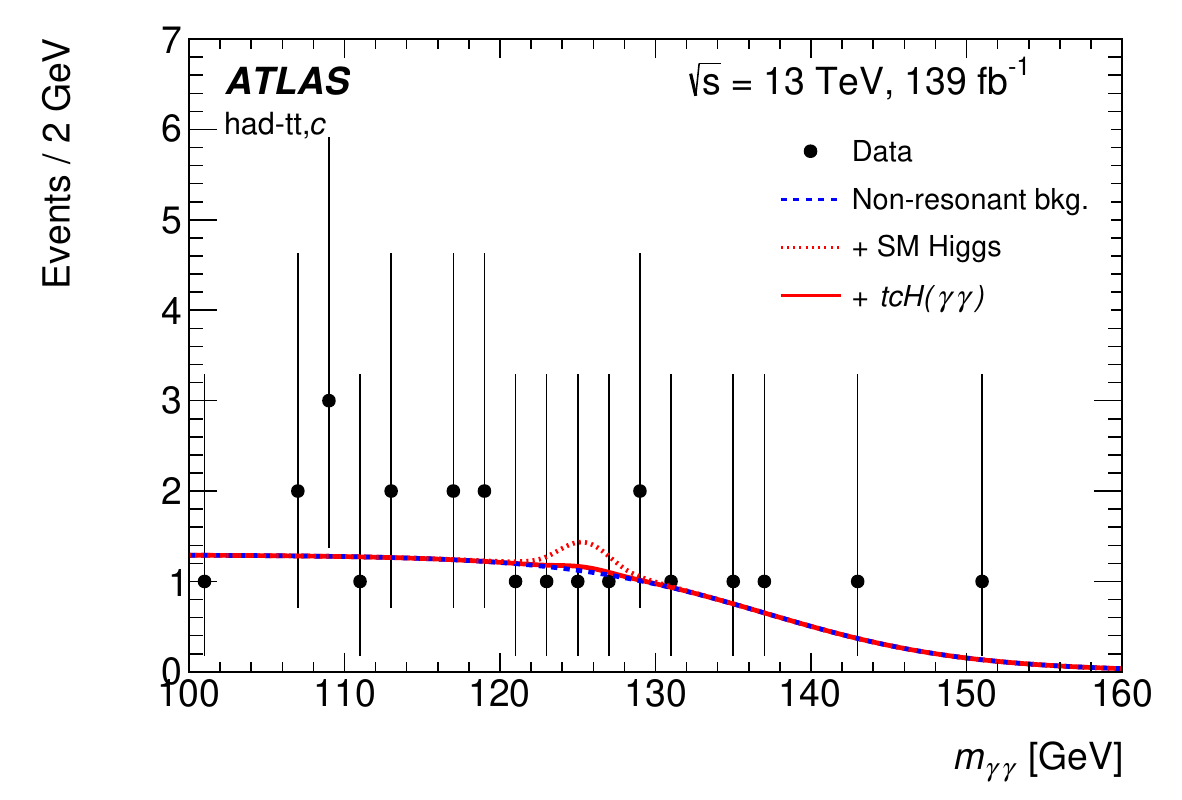}}
\subfigure[]{\includegraphics[width=.45\textwidth]{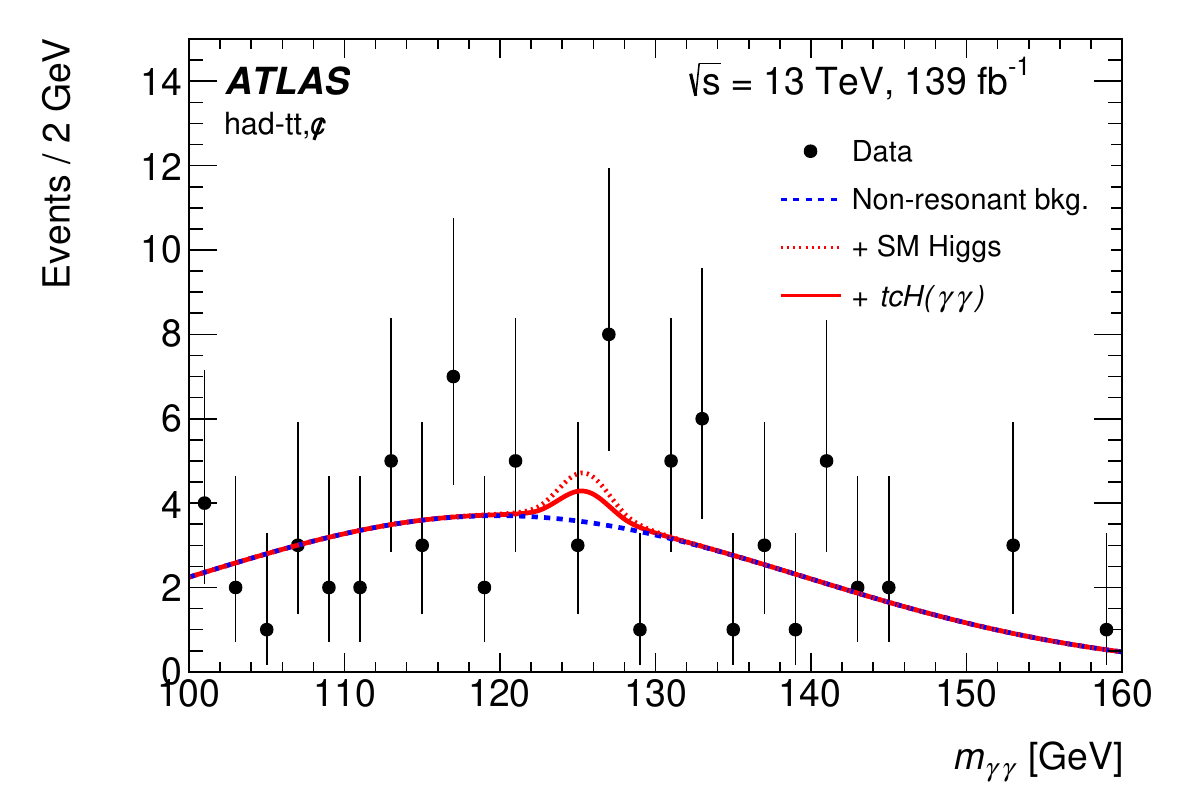}} \\
\subfigure[]{\includegraphics[width=.45\textwidth]{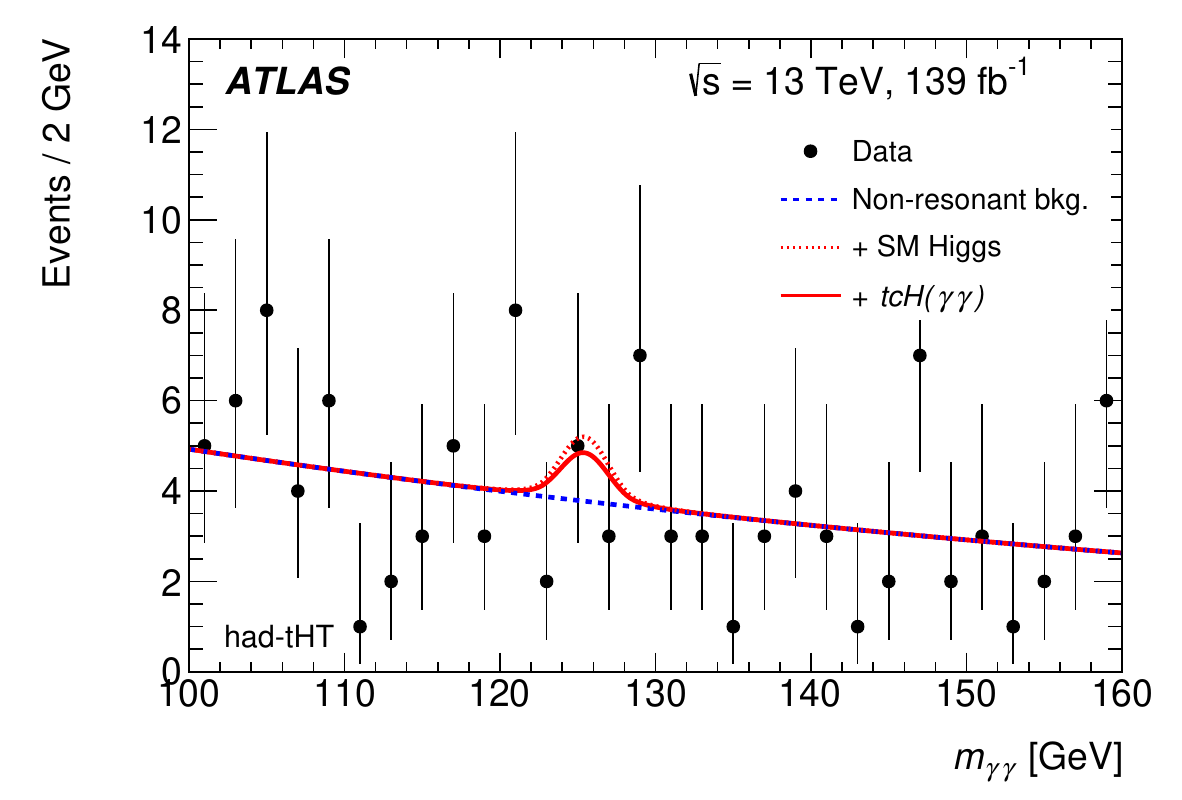}} \\
\subfigure[]{\includegraphics[width=.45\textwidth]{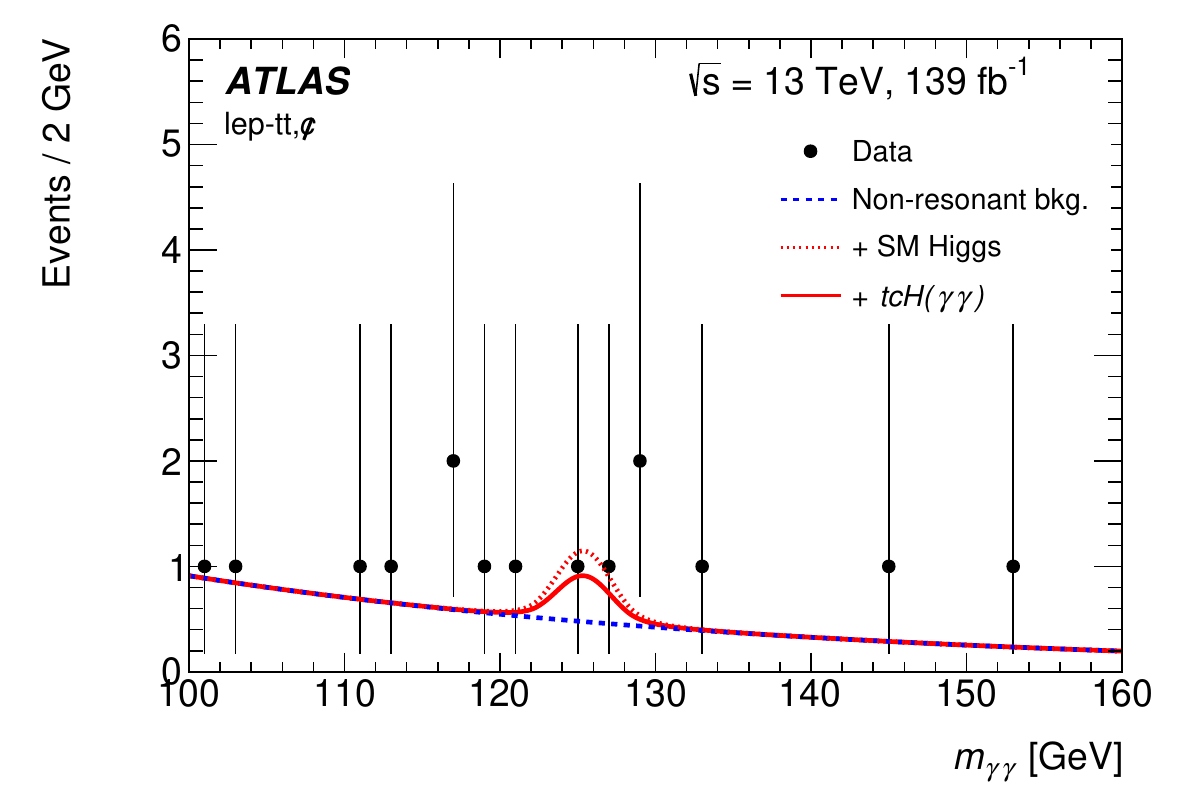}}
\subfigure[]{\includegraphics[width=.45\textwidth]{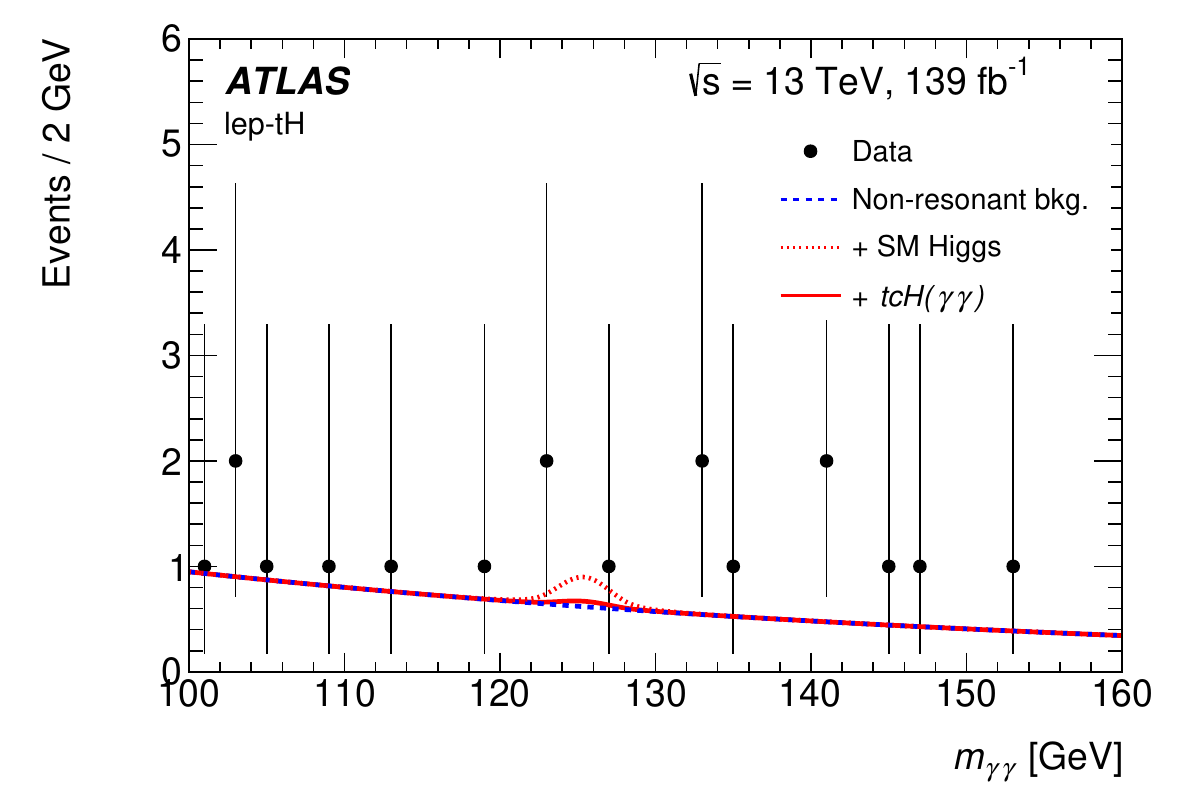}}
\end{center}
\vspace{-0.5cm}
\caption{Distributions of the diphoton invariant mass for the most
relevant categories: three in the hadronic selection ((a) had-\uA,
(b) had-\uB, and (c) had-tHT) and two in the leptonic selection
((d) lep-\uB\ and (e) lep-tH).  The result of a fit to the data of the sum (full line)
of a signal component with the mass of the Higgs boson fixed to $\mH = 125\ \gev$ and assuming a non-zero
$tcH$ coupling, a non-resonant background component (dashed line) and the SM Higgs boson contribution (difference
between the dotted and dashed lines) is superimposed.}
\label{fig:myyShape}
\end{figure}
 
The evolution of $q_{\cal B} = -2\left(\ln L({\cal B})-\ln L({\hat{\cal B}})\right)$ as a function of ${\cal B}$
for each category individually, and for the combined likelihood for the $tcH$ (top) and $tuH$ (bottom) couplings
is presented in figure~\ref{fig:nllvsqB} (left), using an Asimov data set~\cite{Cowan:2010js} assuming the
background-only hypothesis. Several categories contribute in a similar way to the sensitivity to the $tcH$
coupling (had-\uA, had-\uB, lep-\uB, lep-tH), while for $tuH$ the categories had-\uB, lep-\uB, had-tHT and lep-tH dominate.
In figure~\ref{fig:nllvsqB} (right), the evolution of $q_{\cal B}$ for the data is compared with the one obtained
with the Asimov data set.
\begin{figure}[h!]
\begin{center}
\subfigure[]{\includegraphics[width=.49\textwidth]{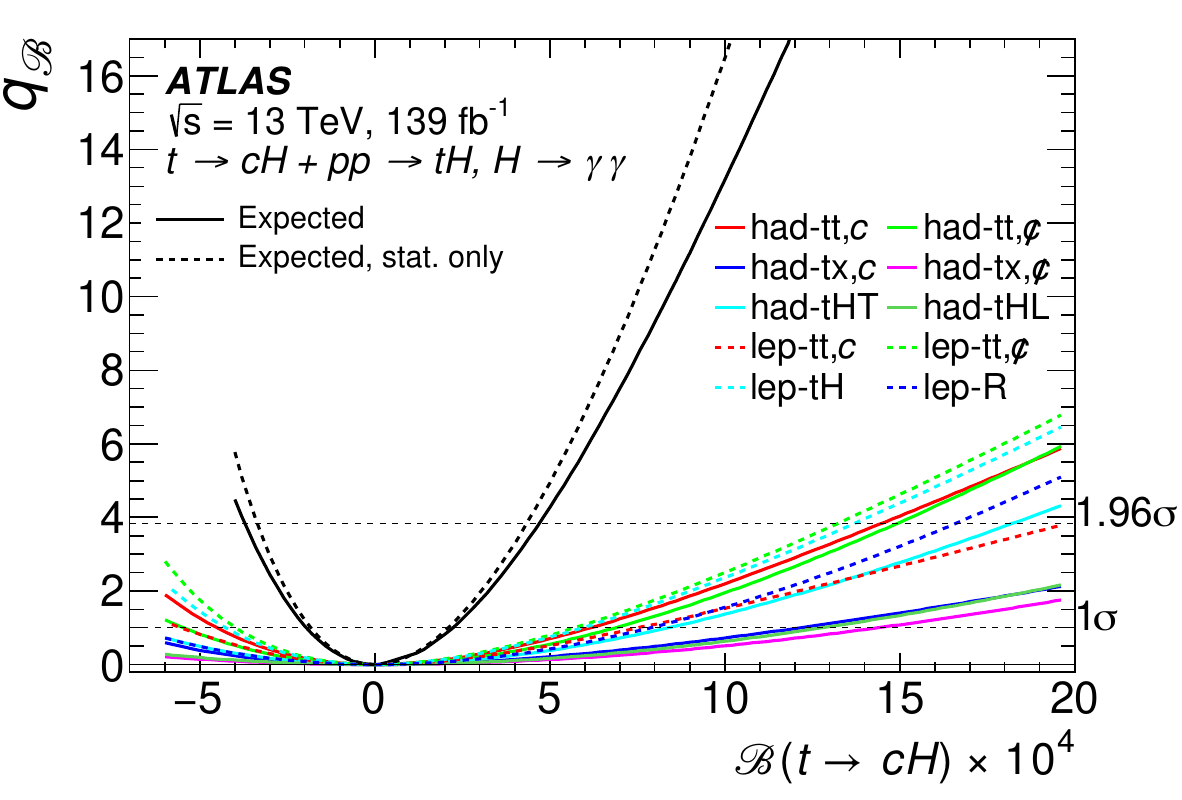}}
\subfigure[]{\includegraphics[width=.49\textwidth]{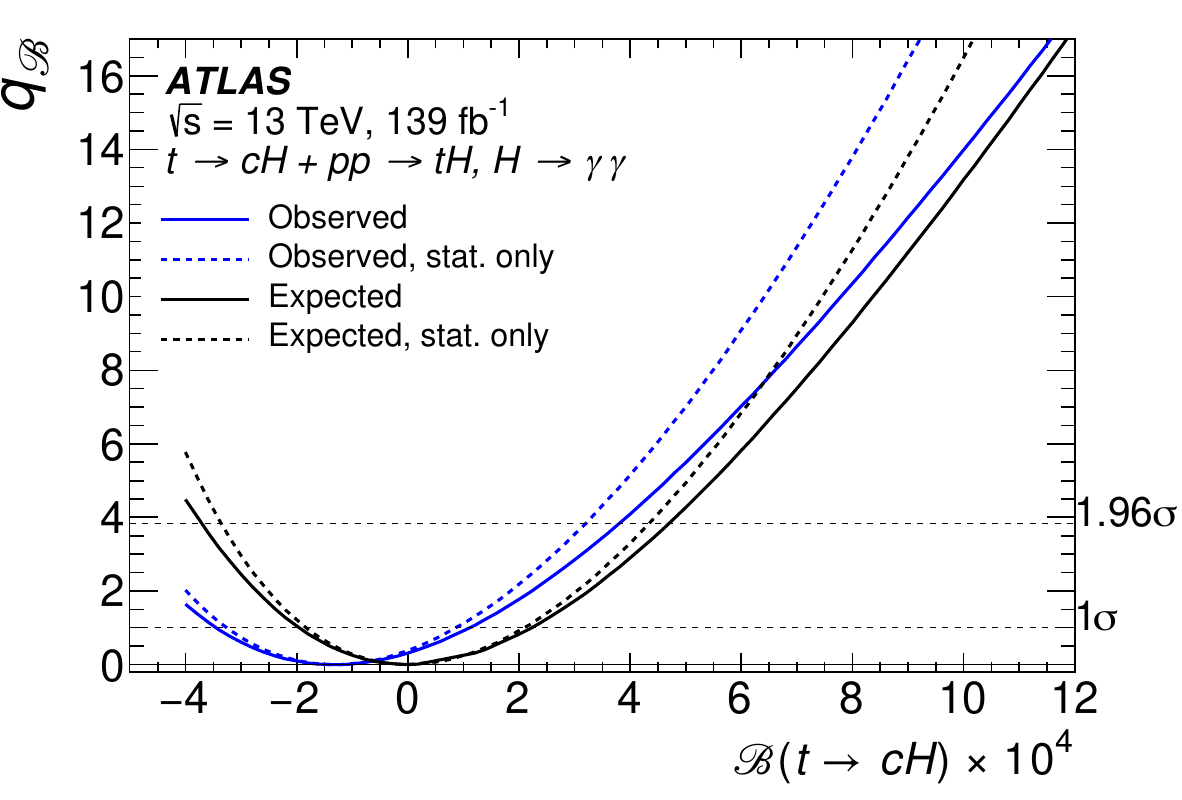}} \\
\subfigure[]{\includegraphics[width=.49\textwidth]{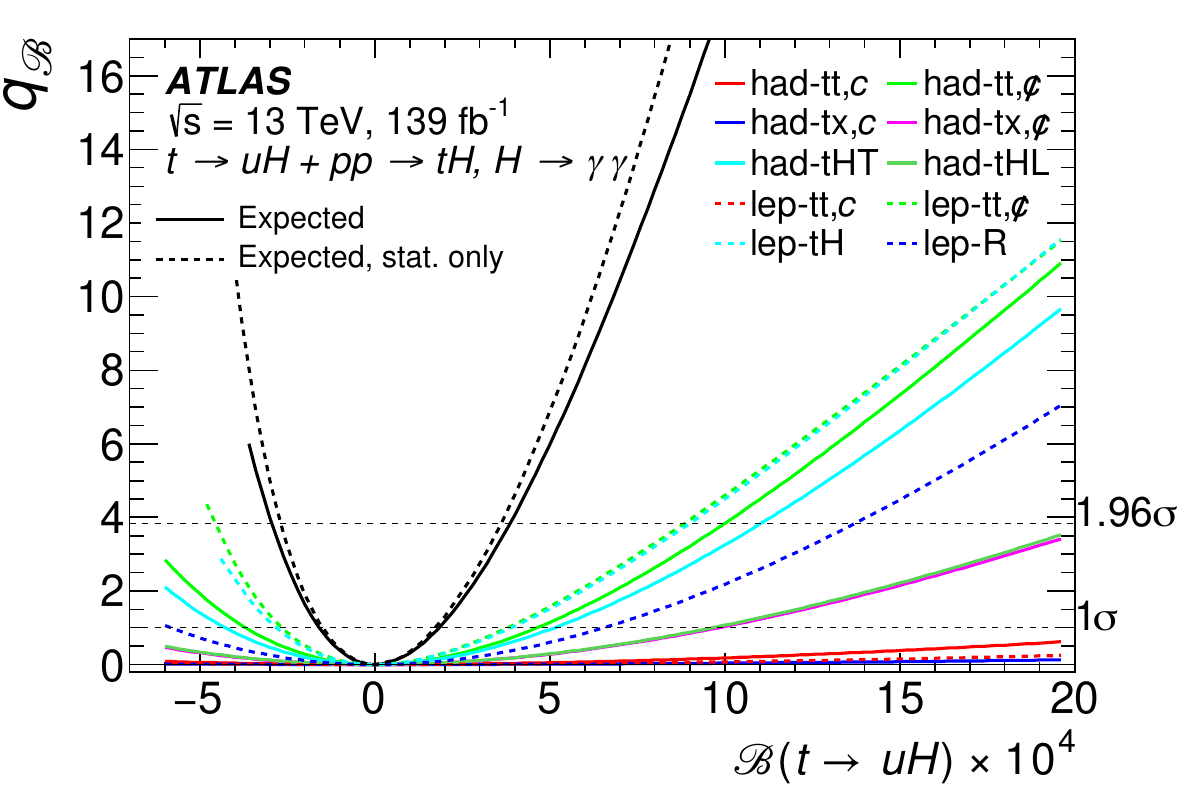}}
\subfigure[]{\includegraphics[width=.49\textwidth]{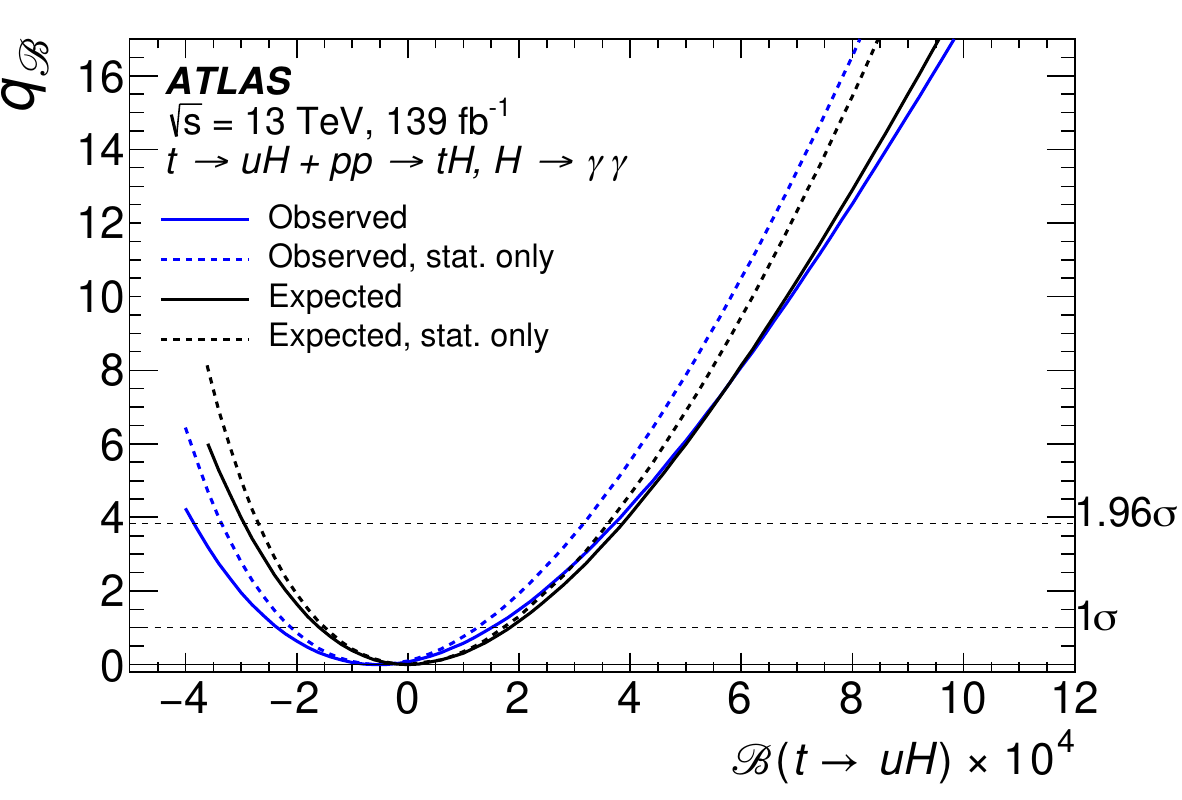}}
\end{center}
\vspace{-0.5cm}
\caption{Evolution of $q_{\cal B}$, the negative log-profile-likelihood ratio
(times 2), as a function of the $t\to cH$ (top) and $t\to uH$ (bottom) branching ratios.
In (a) and (c) the evolutions for each category and the combined sample are shown for the expected result in the absence of signal,
while (b) and (d) show a comparison between the observed result and the expectation in the absence of signal.
The likelihood functions are only defined for a positive expected number of events, hence some categories
and combined curves do not cover the full scanned range.
}
\label{fig:nllvsqB}
\end{figure}
 
The evolution of the signal confidence level, CL$_{\mathrm{s}}$, as a function of the $t \to cH$ branching ratio
is shown in figure~\ref{fig:clsvsbr}. For a confidence level of 5\%, the observed limit on the branching ratio is
$4.3\times10^{-4}$. The expected exclusion limit in the absence of signal is $4.7\times10^{-4}$. For the $tuH$
coupling, the observed limit at 95\% CL is $3.8\times10^{-4}$, while $3.9\times10^{-4}$ is expected in the absence
of signal.
\begin{figure}[h!]
\begin{center}
\includegraphics[width=.49\textwidth]{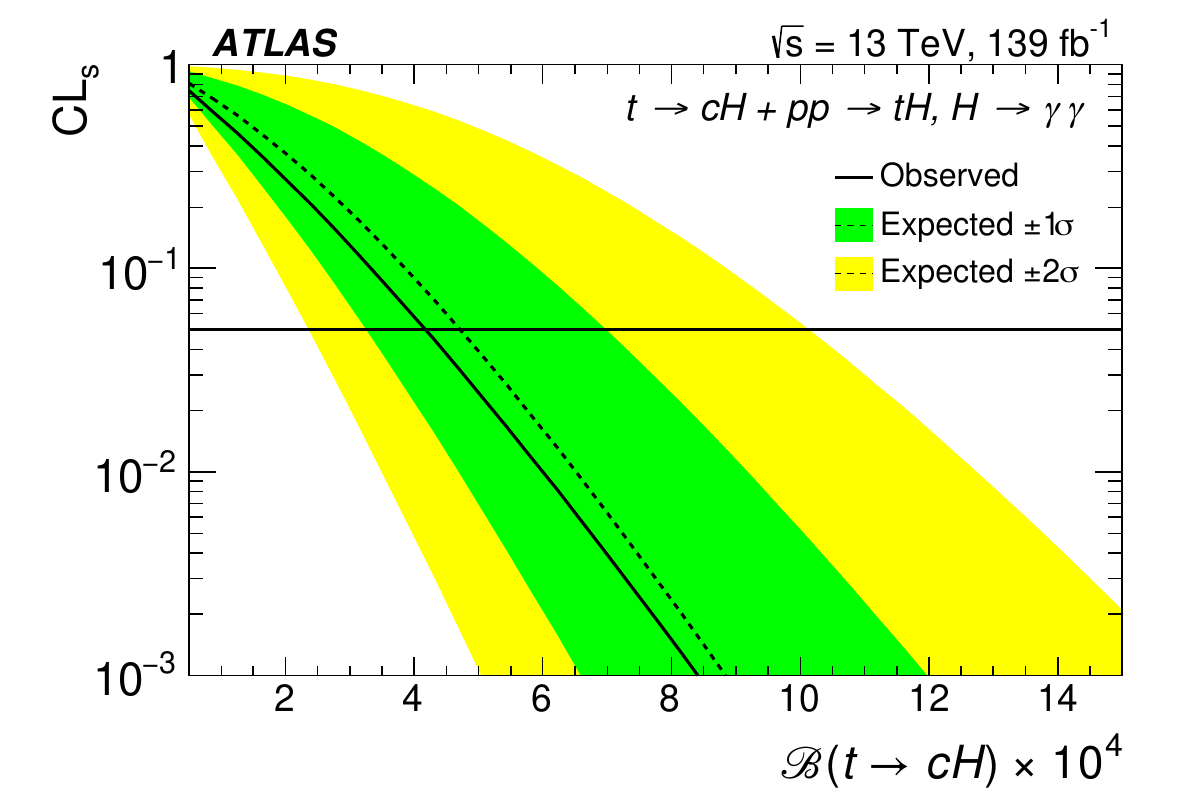}
\end{center}
\vspace{-0.5cm}
\caption{Evolution of the signal confidence level, CL$_{\mathrm{s}}$, as a function of the $t \to cH$ branching ratio ${\cal B}$
for the observation (full line) and the expectation in the absence of signal (dashed line). The bands at $\pm 1 \sigma$ and
$\pm 2 \sigma$ around the expected curve are also shown.}
\label{fig:clsvsbr}
\end{figure}
 
The sensitivity of the analysis is limited by the statistical precision. With all constrained nuisance parameters fixed at
their best fit estimates, the expected 95\% CL upper limit on ${\cal B}$ is $4.4\times10^{-4}$ for the $tcH$
coupling and $3.6\times10^{-4}$ for the $tuH$ coupling.
The most relevant systematic uncertainties affecting the limit on the $tqH$ coupling are presented in table~\ref{tab:sysImp},
where the impact is given as the relative variation of the expected upper limit when fixing the corresponding nuisance
parameters at their best-fit value. The largest contributions to the uncertainty originate from the photon energy resolution,
the \nttb\ cross-section, the $H\to\gamma\gamma$ branching ratio, the parton shower description and the non-resonant
background shape choices.
\begin{table}[h!]
\begin{center}
\caption{Most relevant systematic uncertainties and their relative
impact (in percent) on the expected upper limit on the $t \to qH$ branching ratio.}
\begin{tabular}{lc}
\hline
Source  & relative impact (\%) \\ \hline
Experimental & \\ \hline
Photon energy resolution & 1.5 \\
Photon identification & 0.4 \\
Luminosity, \pileup modelling & 0.3 \\
Jet energy scale and resolution, flavour tagging & $< 0.2$ \\ \hline
Theoretical &  \\ \hline
Normalisation $\left( \sigma(pp\to\ttbar, tH), {\cal B}(H \to
\gamma\gamma) \right)$ & 1.1 \\
Parton showering model & 0.8 \\
$m_{t}$ value, NLO generator for $pp \to tH$ & 0.5 \\
Resonant background & 0.5 \\ \hline
Non-resonant background & 2.3 \\
\hline
\end{tabular}
\label{tab:sysImp}
\end{center}
\end{table}
 
The relevance of the \ctag categorisation has been appraised by studying the separation between the $tcH$ and the $tuH$
hypotheses. The test statistic used to quantify this separation is defined as
$q_c \equiv \left( \ln L(c,{\hat{\hat{\cal B}}}_c)-\ln L(u,{\hat{\hat{\cal B}}}_u) \right)$, where ${\hat{\hat{\cal B}}}_{c(u)}$ is the conditional
maximum-likelihood estimator of ${\cal B}_{c(u)}$ under the hypothesis of a non-zero $tcH$ ($tuH$) coupling.
Distributions of the test statistic for the $tcH$ and $tuH$ hypotheses, for the nominal analysis and the alternative one,
where no \ctag is used, have been built for a $t \to qH$ branching ratio of $10^{-3}$, using $3500$ pseudo-experiments.~\footnote{This
value of the branching ratio has been chosen as it is slightly below the ATLAS combined limit obtained with 36 \ifb~\cite{TOPQ-2017-07},
and would have resulted in a sufficient sensitivity for an observation.} The fractions of experiments from the $tuH$
hypothesis above the median $q_c$ of the distribution for the $tcH$ hypothesis are about 8\% and 40\% for the
nominal and alternative analyses, respectively. Even though the \ctag categorisation has a small impact on the
expected upper limit on the $t\to qH$ branching ratio, it thus has a significant power for the characterisation of a potential signal.
 
\subsection{Combination of ATLAS searches}
The analysis described above has been combined with the corresponding ATLAS searches using the $H \to \tau^+\tau^-$~\cite{TOPQ-2019-17}
and $H \to \bbbar$~\cite{HDBS-2020-16} decays. The correlations between the uncertainties in the different channels have been assessed.
The luminosity uncertainty, the uncertainty in the \pileup modelling and
the uncertainties related to the jet energy scale and energy resolution are correlated among the three channels.
The uncertainties pertaining to \btag are correlated between the $H \to \tau^+\tau^-$ and $H \to \bbbar$ analyses.
The remaining uncertainties (mostly from experimental sources, and signal and background modelling) are taken as uncorrelated.
Although some of the sources of systematic uncertainties (especially related to \btag, electron and muon identification and signal modelling)
are common to the three search channels, they have not been correlated for simplicity.
In each search, the dominant systematic uncertainties are different. In addition, the $H \to \tau^+\tau^-$
and $H \to \gamma\gamma$ channels are dominated by the data's statistical uncertainty.
Therefore, the combined result has a very low sensitivity to the assumed correlations of uncertainties across channels.
 
The observed (expected) 95\% CL combined upper limits on the branching ratios are $5.8\times10^{-4}\ (3.0 \times 10^{-4})$
and $4.0\times 10^{-4}\ (2.4 \times 10^{-4})$ for the decays $t\to cH$ and $t \to uH$, respectively. A summary of the upper limits on the branching ratios
obtained by the individual searches, and their combination, is given in figure~\ref{fig:combLim}.
Using eq.(\ref{eq:yukbr}), the observed upper limit in the $tcH$ combined analysis
corresponds to  $\lambda_{tqH} <0.045$ at 95\%~CL, slightly smaller than the value $\lambda_{tqH}^\text{CS} = 0.057$ obtained by applying
the Cheng and Sher ansatz~\cite{cheng:1987rs}, using running charm-quark and top-quark masses at the top-quark mass of 0.61 \gev\ and 163 \gev,
respectively. In the SMEFT framework, the limit observed for the $t \to cH\ (t \to uH)$ branching ratio translates to a limit on the
corresponding Wilson coefficient of $C^{23,32}_{u\varphi} = 1.07\ (C^{13,31}_{u\varphi} = 0.88)$ at 95\%~CL, assuming
$C^{32,23}_{u\varphi} = 0\ (C^{31,13}_{u\varphi} = 0)$,
and for a mass scale $\Lambda = 1\ \tev$.
\begin{figure}[h!]
\begin{center}
\subfigure[]{\includegraphics[width=.49\textwidth]{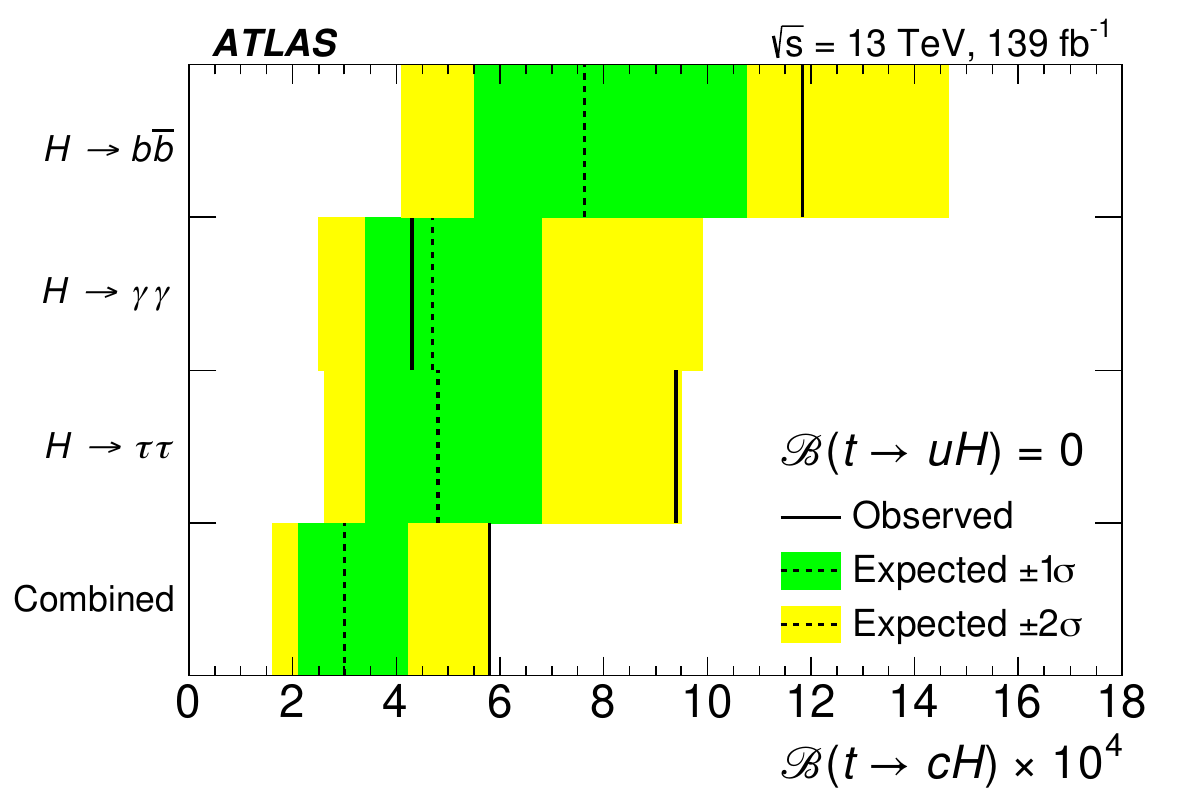}}
\subfigure[]{\includegraphics[width=.49\textwidth]{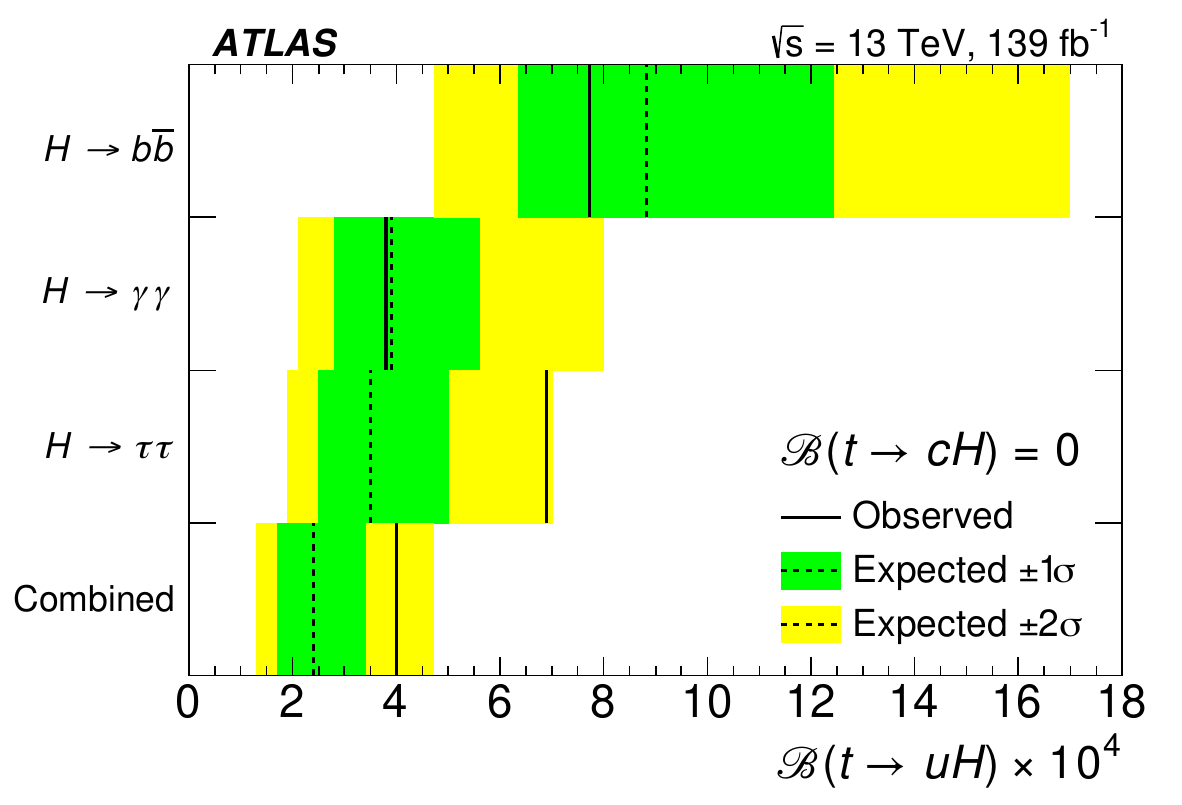}}
\end{center}
\vspace{-0.5cm}
\caption{95\% CL upper limits on  (a) ${\cal B}(t \to cH)$ assuming
${\cal B}(t \to uH) = 0$ and (b) ${\cal B}(t \to uH)$ assuming ${\cal B}(t \to cH) = 0$ for the individual
searches and their combination.
The observed limits (solid lines) are compared with the expected (median) limits under the background-only
hypothesis (dotted lines). The surrounding shaded bands correspond to the 68\% and 95\% CL intervals
around the expected limits, denoted by $\pm 1\sigma$ and $\pm 2 \sigma$, respectively.}
\label{fig:combLim}
\end{figure}

% End of text imported from the .//Sections/Results.tex input file

\section{Conclusion}

% The next lines are included from the .//Sections/Conclu.tex input file
The FCNC coupling of a top quark with a lighter up-type quark $q$ ($q=c,u$) and a Higgs boson
has been searched for in a data set  of 139 \ifb\ of 13 \tev\ proton--proton collisions recorded
by the ATLAS experiment at the LHC, using the Higgs boson decay mode $H \to \gamma\gamma$.
Both the $pp \to \nttb$ production, followed by the top-quark decay $t \to qH$, and the $pp \to tH$ production
are considered in the analysis.
The cross-sections of both processes are parameterized in terms of the $tqH$ coupling strength,
and results are reported as constraints on the branching ratio ${\cal B}$ of the $t \to qH$ decay.
The analysis is split into hadronic
and leptonic channels, targeting the decay of the $W$ boson from the SM top-quark decay either in
a hadronic mode or in a leptonic mode. For each channel the analysis is further split into sub-channels
targeting either $pp \to tH$ or $\pp \to \nttb$ production. In the latter case charm tagging is used
in view of further separating the $tcH$ coupling from the $tuH$ one.
 
Exploiting the diphoton invariant mass distributions, a sideband technique is used to constrain
the background under the signal. Taking into account the contribution of the SM Higgs boson
production, an upper limit on the $t \to cH\ (t \to uH)$ decay branching ratio
in the absence of signal of ${\cal B} = 4.7\times 10^{-4}$ ($3.9\times 10^{-4}$) is expected at 95\% CL.
For the $tcH$ coupling, the sensitivity increase relative to the analysis described in ref.~\cite{HIGG-2016-26}
is a factor of $\sim 1.5$ better than expected from the increase of the integrated luminosity alone,
thanks to an improved event reconstruction and categorisation. No statistically significant excess is observed in the data,
and a limit on ${\cal B}$ of $4.3\times 10^{-4}$ ($3.8\times 10^{-4}$) is set at the 95\%  CL for $\mH = 125\ \gev$.
 
The combination of this search with ATLAS searches using the Higgs boson decays $H\to\tau^+\tau^-$ and $H\to\bbbar$
yields observed (expected) 95\% CL upper limits on the $t \to cH$ branching ratio
of $5.8\times10^{-4}\ (3.0 \times 10^{-4})$ assuming ${\cal B}(t \to uH) = 0$, and on the $t \to uH$ branching ratio of $4.0 \times 10^{-4}\ (2.4 \times 10^{-4})$
assuming ${\cal B}(t \to cH) = 0$. Corresponding limits on the relevant Wilson
coefficients in the SMEFT framework are also derived.

% End of text imported from the .//Sections/Conclu.tex input file

\section*{Acknowledgements}

% The next lines are included from the .//acknowledgements/Acknowledgements.tex input file

We thank CERN for the very successful operation of the LHC, as well as the
support staff from our institutions without whom ATLAS could not be
operated efficiently.
 
We acknowledge the support of
ANPCyT, Argentina;
YerPhI, Armenia;
ARC, Australia;
BMWFW and FWF, Austria;
ANAS, Azerbaijan;
CNPq and FAPESP, Brazil;
NSERC, NRC and CFI, Canada;
CERN;
ANID, Chile;
CAS, MOST and NSFC, China;
Minciencias, Colombia;
MEYS CR, Czech Republic;
DNRF and DNSRC, Denmark;
IN2P3-CNRS and CEA-DRF/IRFU, France;
SRNSFG, Georgia;
BMBF, HGF and MPG, Germany;
GSRI, Greece;
RGC and Hong Kong SAR, China;
ISF and Benoziyo Center, Israel;
INFN, Italy;
MEXT and JSPS, Japan;
CNRST, Morocco;
NWO, Netherlands;
RCN, Norway;
MEiN, Poland;
FCT, Portugal;
MNE/IFA, Romania;
MESTD, Serbia;
MSSR, Slovakia;
ARRS and MIZ\v{S}, Slovenia;
DSI/NRF, South Africa;
MICINN, Spain;
SRC and Wallenberg Foundation, Sweden;
SERI, SNSF and Cantons of Bern and Geneva, Switzerland;
MOST, Taiwan;
TENMAK, T\"urkiye;
STFC, United Kingdom;
DOE and NSF, United States of America.
In addition, individual groups and members have received support from
BCKDF, CANARIE, Compute Canada and CRC, Canada;
PRIMUS 21/SCI/017 and UNCE SCI/013, Czech Republic;
COST, ERC, ERDF, Horizon 2020, ICSC-NextGenerationEU and Marie Sk{\l}odowska-Curie Actions, European Union;
Investissements d'Avenir Labex, Investissements d'Avenir Idex and ANR, France;
DFG and AvH Foundation, Germany;
Herakleitos, Thales and Aristeia programmes co-financed by EU-ESF and the Greek NSRF, Greece;
BSF-NSF and MINERVA, Israel;
Norwegian Financial Mechanism 2014-2021, Norway;
NCN and NAWA, Poland;
La Caixa Banking Foundation, CERCA Programme Generalitat de Catalunya and PROMETEO and GenT Programmes Generalitat Valenciana, Spain;
G\"{o}ran Gustafssons Stiftelse, Sweden;
The Royal Society and Leverhulme Trust, United Kingdom.
 
The crucial computing support from all WLCG partners is acknowledged gratefully, in particular from CERN, the ATLAS Tier-1 facilities at TRIUMF (Canada), NDGF (Denmark, Norway, Sweden), CC-IN2P3 (France), KIT/GridKA (Germany), INFN-CNAF (Italy), NL-T1 (Netherlands), PIC (Spain), ASGC (Taiwan), RAL (UK) and BNL (USA), the Tier-2 facilities worldwide and large non-WLCG resource providers. Major contributors of computing resources are listed in Ref.~\cite{ATL-SOFT-PUB-2023-001}.

% End of text imported from the .//acknowledgements/Acknowledgements.tex input file

\clearpage

\printbibliography
 
\clearpage
% ATLAS Collaboration author list
% Reference date of TOPQ-2019-04 is 2023-06-21
% Author list last updated on date 12-JAN-24
% Data extracted on 12-Jan-2024 for paper reference TOPQ-2019-04
% at 10:48am
 
\begin{flushleft}
\hypersetup{urlcolor=black}
{\Large The ATLAS Collaboration}

\bigskip

\AtlasOrcid[0000-0002-6665-4934]{G.~Aad}$^\textrm{\scriptsize 102}$,
\AtlasOrcid[0000-0002-5888-2734]{B.~Abbott}$^\textrm{\scriptsize 120}$,
\AtlasOrcid[0000-0002-1002-1652]{K.~Abeling}$^\textrm{\scriptsize 55}$,
\AtlasOrcid[0000-0001-5763-2760]{N.J.~Abicht}$^\textrm{\scriptsize 49}$,
\AtlasOrcid[0000-0002-8496-9294]{S.H.~Abidi}$^\textrm{\scriptsize 29}$,
\AtlasOrcid[0000-0002-9987-2292]{A.~Aboulhorma}$^\textrm{\scriptsize 35e}$,
\AtlasOrcid[0000-0001-5329-6640]{H.~Abramowicz}$^\textrm{\scriptsize 151}$,
\AtlasOrcid[0000-0002-1599-2896]{H.~Abreu}$^\textrm{\scriptsize 150}$,
\AtlasOrcid[0000-0003-0403-3697]{Y.~Abulaiti}$^\textrm{\scriptsize 117}$,
\AtlasOrcid[0000-0002-8588-9157]{B.S.~Acharya}$^\textrm{\scriptsize 69a,69b,m}$
% ...
\bigskip
\\

% ...
$^{29}$Physics Department, Brookhaven National Laboratory, Upton NY; United States of America.\\
% ...
$^{35}$$^{(a)}$Facult\'e des Sciences Ain Chock, R\'eseau Universitaire de Physique des Hautes Energies - Universit\'e Hassan II, Casablanca;$^{(b)}$Facult\'{e} des Sciences, Universit\'{e} Ibn-Tofail, K\'{e}nitra;$^{(c)}$Facult\'e des Sciences Semlalia, Universit\'e Cadi Ayyad, LPHEA-Marrakech;$^{(d)}$LPMR, Facult\'e des Sciences, Universit\'e Mohamed Premier, Oujda;$^{(e)}$Facult\'e des sciences, Universit\'e Mohammed V, Rabat;$^{(f)}$Institute of Applied Physics, Mohammed VI Polytechnic University, Ben Guerir; Morocco.\\
% ...
$^{49}$Fakult\"{a}t Physik , Technische Universit{\"a}t Dortmund, Dortmund; Germany.\\
% ...
$^{55}$II. Physikalisches Institut, Georg-August-Universit\"{a}t G\"ottingen, G\"ottingen; Germany.\\
% ...
$^{69}$$^{(a)}$INFN Gruppo Collegato di Udine, Sezione di Trieste, Udine;$^{(b)}$ICTP, Trieste;$^{(c)}$Dipartimento Politecnico di Ingegneria e Architettura, Universit\`a di Udine, Udine; Italy.\\
% ...
$^{102}$CPPM, Aix-Marseille Universit\'e, CNRS/IN2P3, Marseille; France.\\
% ...
$^{117}$Department of Physics, New York University, New York NY; United States of America.\\
% ...
$^{120}$Homer L. Dodge Department of Physics and Astronomy, University of Oklahoma, Norman OK; United States of America.\\
% ...
$^{150}$Department of Physics, Technion, Israel Institute of Technology, Haifa; Israel.\\
$^{151}$Raymond and Beverly Sackler School of Physics and Astronomy, Tel Aviv University, Tel Aviv; Israel.\\
% ...
$^{a}$ Also Affiliated with an institute covered by a cooperation agreement with CERN.\\
$^{b}$ Also at An-Najah National University, Nablus; Palestine.\\
$^{c}$ Also at Borough of Manhattan Community College, City University of New York, New York NY; United States of America.\\
$^{d}$ Also at Center for High Energy Physics, Peking University; China.\\
$^{e}$ Also at Center for Interdisciplinary Research and Innovation (CIRI-AUTH), Thessaloniki; Greece.\\
$^{f}$ Also at Centro Studi e Ricerche Enrico Fermi; Italy.\\
% ...
$^{m}$ Also at Department of Physics, King's College London, London; United Kingdom.\\
% ...

\end{flushleft}

% Created with Glance <Atlas.Glance@cern.ch>

\end{document}